\documentclass[prb,floatfix,twocolumn,superscriptaddress,showpacs,amsmath,amssymb,showpacs,longbibliography]{revtex4-2}

\usepackage{graphicx}
\usepackage{epstopdf}
\usepackage{epsfig}
\usepackage{epsf}
\usepackage{url}
\usepackage[USenglish]{babel}
\usepackage{hyperref}
\def\bcen{\begin{center}}
\def\ecen{\end{center}}
\allowdisplaybreaks
\renewcommand\[{\begin{equation}}
\renewcommand\]{\end{equation}}
\usepackage{verbatim}
\usepackage{natbib}
\usepackage{amsmath, nccmath}
\usepackage{dcolumn}
\usepackage{bm}
\usepackage{bbm}
\usepackage{lipsum}

\usepackage{orcidlink}

\begin{document}
\title{Self-consistent $GW$ + Extended Dynamical Mean Field Theory for semiconductors and insulators}
\author{Viktor Christiansson\orcidlink{0000-0002-9002-0827}}
\affiliation{Department of Physics, University of Fribourg, 1700 Fribourg, Switzerland}
\author{Francesco Petocchi\orcidlink{0000-0002-8580-781X}}
\affiliation{Department of Physics, University of Fribourg, 1700 Fribourg, Switzerland}
\affiliation{Department of Quantum Matter Physics, University of Geneva, 1211 Geneva 4, Switzerland}
\author{Philipp Werner\orcidlink{0000-0002-2136-6568}}
\affiliation{Department of Physics, University of Fribourg, 1700 Fribourg, Switzerland}

\begin{abstract}
Theoretical studies of semiconductors and band insulators are usually based on variants of the $GW$ method without full self-consistency, like single-shot $G^0W^0$ or quasiparticle self-consistent $GW$.
Fully self-consistent $GW$ provides a poor description of the gap size and electronic structure
due to the lack of vertex corrections. While it is hard to predict at which order corrections can be neglected, local vertex corrections to all orders can be consistently included by combining $GW$ with extended dynamical mean field theory (EDMFT).
Here, we show that \textit{ab initio} multitier 
$GW$+EDMFT calculations, which achieve full self-consistency in a suitably defined low-energy space, provide
an accurate description of semiconductors and band insulators, comparable to the established methods which are typically used for this class of materials. Our results demonstrate that 
the range of applicability of $GW$+EDMFT extends to weakly correlated systems,
and they imply that despite the weak correlations,
local vertex corrections 
are an important ingredient in the diagrammatic 
treatment of semiconductors and band insulators. 
\end{abstract}

\maketitle

\section{Introduction} Semiconductors and band insulators are intensely studied materials with technological importance.
The established theoretical description of these systems is based either on density functional theory (DFT) \cite{Hohenberg1964,Kohn1965,Sham1983,Xiao2011} or various flavors of the $GW$ approximation \cite{Hedin1965} to the electronic self-energy.
In the latter case, the calculations are often performed in a one-shot fashion, dubbed $G^0W^0$, where a non-interacting Green's function $G^0$ (typically constructed from the Kohn-Sham energies and wavefunctions of a DFT calculation) is used to obtain the self-energy $\Sigma$. Already early calculations showed that this approach produces results in generally good agreement with experiment \cite{Strinati1980,Strinati1982,Hybertsen1986}. From a conceptual point of view, it is however not fully satisfactory, because the result depends on the specific DFT input.

Quasiparticle self-consistent $GW$ (QSGW) \cite{Schilfgaarde2006,Kotani2007}, which implements a partial self-consistency and thus removes some of the arbitrariness in the starting $G^0$, was an important step forward in this respect. Although QSGW still uses a noninteracting Green's function $G^0$ constructed from a quasiparticle Hamiltonian, the approach leads to improved estimates of the band gaps compared to experiment. 
Other partially self-consistent schemes, such as eigenvalue self-consistency, $GW^0$ where the screened interaction $W^0$ is kept fixed, or methods which only update the Hartree and exchange parts in the self-consistent cycle, have also been proposed \cite{Northrup1987,Luo2002,Shishkin2007,Stan2009,Hellgren2021}.
Fully self-consistent $GW$ (sc$GW$) calculations, on the other hand, are rarely used, since they lack relevant cancellation effects \cite{Gukelberger2015}, and 
in many cases significantly overestimate the band gaps \cite{Kutepov2012,Kutepov2017}.

In terms of diagrams, $GW$ corresponds to a lowest order approximation of the self-energy in $W$.
A good agreement with experiment could be ``accidental", in the sense that it relies on cancellations between higher order vertex corrections, and in the case of non-self-consistency be due to a suitable starting point.
(QSGW features an approximate cancellation inherent to the method, which can be understood from a  ``Z factor cancellation" between the lack of vertex corrections and the quasi-particle approximation, at least when the dominant interaction contribution is long-ranged \cite{Kotani2007,Kutepov2016}.)
$GW$ is furthermore known to suffer from problems with a physical origin, such as the 
inherent self-screening error \cite{Nelson2007,Aryasetiawan2012}, which for several semiconductors leads to inaccuracies in the calculated band gap comparable in magnitude to the difference between $G^0W^0$ and experiment \cite{Christiansson2023a}. 
Kutepov showed that including vertex corrections to the first order beyond $GW$ (vc$GW$) \cite{Kutepov2017} has a significant effect in semiconductors and band insulators, and in many cases leads to improvements compared to conventional sc$GW$ and QSGW results.
Similar vertex corrections were built into QSGW in Ref.~\cite{Cunningham2023}, and also resulted in a better agreement with experiment. An earlier work by Gr\"uneis \emph{et al.} \cite{Gruneis2014} included vertex corrections with a static $W$ and similarly concluded that an improved agreement with experimental properties can be achieved when vertex corrections are considered.
It is, however, difficult to know {\it a priori} at which point it is possible to neglect further corrections. 

A method which allows us to clarify the role of local vertex corrections is the $GW$+extended dynamical mean-field theory ($GW$+EDMFT) approach \cite{Biermann2003,Boehnke2016,Nilsson2017}, which combines an EDMFT treatment of the local correlations and screening to all orders with the nonlocal contributions from $GW$. The scheme has no double-counting issues, 
no adjustable parameters apart from a separation into high- and low-energy subspaces, and it implements a full self-consistency in the low-energy space. 
\textit{Ab initio} $GW$+EDMFT has so far been mostly applied to moderately-to-strongly correlated materials such as cubic perovskites \cite{Boehnke2016,Petocchi2020a,Kang2024,Mushkaev2024}, ruthenates \cite{Petocchi2021}, and nickelates \cite{Petocchi2020b,Christiansson2023b,Christiansson2023c}. Calculations for more weakly correlated metallic systems have shown mixed results \cite{Nilsson2017,Christiansson2022a}, and tests on bulk semiconductors have been limited to a one-shot calculation for Si \cite{Zhu2021}.
To assess the predictive power of $GW$+EDMFT, it is thus important to perform systematic tests of the full scheme on weakly correlated compounds, and to compare the results to established methods.

Here, we study elemental and binary semiconductors using {\it ab initio} multitier $GW$+EDMFT \cite{Boehnke2016,Nilsson2017}. The relevant states that we treat self-consistently in a low-energy window around the band gap can be described by a model consisting of the $s$ and $p$ states of the diatomic primitive unit cell, i.e. an 8-orbital low-energy model. (For systems with shallow semicore $d$ states we also include these in the low-energy model.) In the EDMFT calculation, we solve a separate impurity problem involving these $sp$ states for each of the atomic sites, and thus treat intersite correlations only at the $GW$ level.
For most of the considered materials, we find that $GW$+EDMFT
yields band gaps comparable to, or closer to experiment than one-shot or quasiparticle self-consistent $GW$, and that the agreement with experiment is on par with other vertex corrected schemes. We also show that the method leads to improvements in the position of the semicore $d$ states for several materials, and that it gives a good prediction of photoemission spectra. 

\begin{figure}[t!] 
\begin{centering}
\includegraphics[width=1.0\columnwidth]{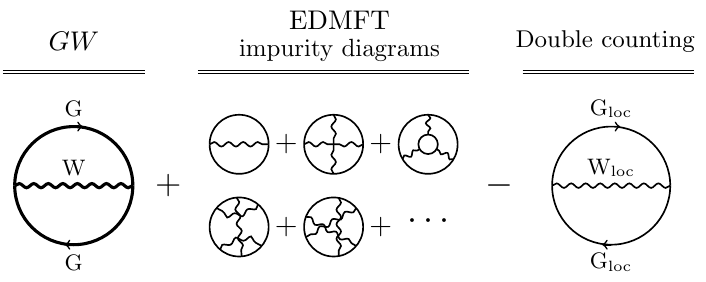}
\par\end{centering}
\caption{
\label{Fig:diagrams} Diagrammatic representation of the terms contributing to the $\Psi^{GW+\mathrm{EDMFT}}$ 
functional, with prefactors omitted, following Ref.~\cite{Nilsson2017}. The labels for the local $G_\textrm{imp}$ (thin straight lines) and $W_\textrm{imp}$ (thin wavy lines) in the EDMFT diagrams are suppressed for better readability. These lines correspond to the impurity Green's function and screened interaction, respectively. Thick straight and wavy lines correspond to lattice propagators. The double counting term is the local projection of the $GW$ diagram (see text).
}
\end{figure}

\section{Method} 

\subsection{$GW$+EDMFT}

In $GW$+EDMFT the lowest-order $GW$ self-energy and polarization functions are supplemented with all local diagrams (vertex corrections) from an EDMFT impurity calculation \cite{Biermann2003,Boehnke2016}.
Following Ref.~\cite{Biermann2003}, the $GW$+EDMFT method can be shown to be a $\Psi$-derivable theory.
We start from the free-energy functional \cite{ALMBLADH1999,Chitra2001}
\begin{align}
 \Gamma[G,W] &= \mathrm{Tr} \ln G - \mathrm{Tr} \left[ \left( G_H^{-1}  - G^{-1} \right) G  \right] \\
&-\frac{1}{2}\mathrm{Tr} \ln W
+\frac{1}{2}\mathrm{Tr} \left[ \left( v^{-1}  - W^{-1} \right) W  \right] + \Psi[G,W], \nonumber
\end{align}
where $G_H$ is the non-interacting Hartree Green's function, $v$ the bare Coulomb interaction and $G$ and $W$ the interacting Green's function and dynamically screened interaction, respectively. In this expression, the $\Psi$ functional contains all contributions beyond Hartree \cite{Biermann2003}. The relations between $\Psi$ and the self-energy $\Sigma$ and polarization function $\Pi$ follow directly from the requirement of stationary of $\Gamma$ and a comparison with the fermionic and bosonic Dyson equations:
\begin{equation}
\begin{aligned}
G^{-1} &= G_H^{-1} - \frac{\delta \Psi}{\delta G}    \\
W^{-1} &= v^{-1} + 2\frac{\delta \Psi}{\delta W}   
\end{aligned}
\quad \longrightarrow \quad
\begin{aligned}
\frac{\delta \Psi}{\delta G}&=\Sigma ,\\
\frac{\delta \Psi}{\delta W} &= - \frac{1}{2}\, \Pi .
\end{aligned}
\end{equation}
In this framework, the standard (self-consistent) $GW$ approximation corresponds to \cite{ALMBLADH1999} $\Psi^{GW}[G,W]=-\frac{1}{2}GWG$ with $\Sigma^{GW}=-GW$ and the polarization function given by the (random-phase approximation, RPA) bubble $\Pi^{GW}=GG$.

In the original formulation of the $GW$+EDMFT scheme by Biermann and coworkers \cite{Biermann2003}, the full functional is split into two parts: a fully nonlocal and fully local one. The purely nonlocal contributions are treated on the $GW$ level, whereas the local part is taken from an EDMFT impurity problem \cite{Georges1996,Sun2002}, where (local) diagrams to all orders are included nonperturbatively. Local here means on-site, i.e., between localized orbitals centered on an atomic site.

As is directly evident from the diagrammatic representations of $\Psi^{GW}$ and $\Psi^\mathrm{EDMFT}$ in Fig.~\ref{Fig:diagrams}, the only diagram that is included in both the $GW$ and EDMFT schemes is the (site-)local projection of the $GW$ one. The total functional can therefore be written as the sum of the two, with the double counting removed.
The resulting $\Psi$ functional for $GW$+EDMFT becomes
\begin{align}
\Psi^{GW\mathrm{+EDMFT}}[G,W]  &= \Psi^{GW}[G,W] + \Psi^\mathrm{EDMFT}[G_\mathrm{loc},W_\mathrm{loc}] \nonumber \\ 
&- \Psi^{GW}[G_\mathrm{loc},W_\mathrm{loc}],
\end{align}
where at self-consistency the local projection of $G$ and $W$ must coincide with the ones derived from the quantum impurity problem with all local diagrams $\Psi^\mathrm{EDMFT}$: $G_\textrm{loc}=G_\textrm{imp}$ and $W_\textrm{loc}=W_\textrm{imp}$.

An alternative point of view is to instead directly consider the standard vertex function $\Lambda$ in the Hedin equations \cite{Hedin1965} (here at finite temperature) which can be separated into a trivial part and vertex corrections:
\begin{align}
\Lambda(1,2,3) &= \delta(1-2)\delta(2-3) \\ &+ \int d(4567) \frac{\partial \Sigma(1,2)}{\partial G(4,5)}G(4,6) G(7,5) \Lambda(6,7,3), \nonumber
\end{align}
with $1=(t_1,{\bf r_1})$ and spin indices suppressed.
As a consequence, the full self-energy is split into two parts: $\Sigma=-GW\Lambda=\Sigma^{GW} + \Sigma^\mathrm{vc}$, where the first term is the standard $GW$ approximation (also derived from the $\Psi$ functional above), and the vertex term $\Sigma^\mathrm{vc}$ contains everything beyond. The same procedure also applies to $\Pi$. 

Vertex corrections beyond the $GW$ approximation ($\Sigma^\mathrm{vc}$ and $\Pi^\mathrm{vc}$) can then be explicitly included. Typically, these need to be truncated at a very low order in practical calculations, in particular if the full nonlocal structure is considered, necessitating a judicious choice of diagrams \cite{Kutepov2016}. 
Alternatively, a certain class of vertex corrections can be considered and summed to infinite order, removing the ambiguity of when to truncate the series of diagrams.
The latter approach is taken in $GW$+EDMFT, where the subset of vertex corrections that are fully site-local are retained, with a feedback between the local and nonlocal contributions in both $G$ and the dynamically screened $W$ through the self-consistency loop.
These local vertex corrections to the self-energy and polarization are obtained from the solution of a self-consistently determined effective impurity problem calculated within the EDMFT framework. Crucially, in this approach the local terms are treated nonperturbatively to infinite order. 
For additional details see Refs.~\cite{Biermann2003,Nilsson2017}.

In $GW$+EDMFT, the polarization function $\Pi$ and self-energy $\Sigma$ are calculated as
\begin{align}
\Sigma^{GW+\textrm{EDMFT}}_{\bf k} &= \Sigma^{GW}_{\bf k} + \Sigma^{\textrm{EDMFT}}_\textrm{imp} - \Sigma^\mathrm{DC}, \label{Eq:Sigma} \\
\Pi^{GW+\textrm{EDMFT}}_{\bf q} &= \Pi^{GW}_{\bf q} + \Pi^{\textrm{EDMFT}}_\textrm{imp}  - \Pi^\mathrm{DC} \label{Eq:Pi},
\end{align}
where the double counting terms 
can be derived from the $\Psi[G,W]$ functional presented above, and correspond to the local versions of the $GW$ self-energy and polarization $\left(\Sigma^\mathrm{DC}=-{G}_\textrm{loc}{W}_\textrm{loc}\right.$ and $\left.\Pi^\mathrm{DC}={G}_\textrm{loc}{G}_\textrm{loc}\right)$.
Within the multitier approach, the $GW$ contribution is further split up by defining a low-energy subspace, which encompasses the most relevant lower energy degrees of freedom. 
In the low-energy space, the $GW$+EDMFT equations are solved self-consistently, keeping the full momentum- and frequency-dependence of the functions, as well as the off-diagonal 
nonlocal components \cite{Nilsson2017}. 
In each iteration the screened interaction is updated using the bosonic Dyson equation
\begin{equation}
W^{-1}_{\mathbf q} = v_{\mathbf q}^{-1} - \Pi_{\mathbf q}^{GW+\textrm{EDMFT}},
\end{equation}
with $\Pi$ in Eq.~\eqref{Eq:Pi} calculated using the dressed (interacting) $G$ from the previous iteration. The new $GW$ self-energy, as well as double counting, is then calculated using this screened interaction and the $G$ from the previous iteration (containing the local vertex corrections through the impurity quantities). The new self-energy is combined with the EDMFT (local) self-energy from the previous iteration (Eq.~\eqref{Eq:Sigma}), and the interacting $G$ is recalculated through the fermionic Dyson equation
\begin{equation}
G^{-1}_{\mathbf k} = G^{0\, -1}_{\mathbf k} - \Sigma^{GW+\textrm{EDMFT}}_{\mathbf k}.
\end{equation}
We have here written the self-energy and polarization contributions from the different subspaces (tiers) together for simplicity (see Ref.~\cite{Nilsson2017} for the detailed expressions). 
From the local projection of the dressed quantities, $G_\mathrm{loc}$ and $W_\mathrm{loc}$, one can then calculate new fermionic and bosonic Weiss fields (the dynamical bare propagators $\mathcal{G}$ and $\mathcal{U}$) for the EDMFT impurity problem:
\begin{align}
\mathcal{G}&=\left[\Sigma_\mathrm{imp} + G_\mathrm{loc}^{-1}\right]^{-1}, \\
\mathcal{U}&=W_\mathrm{loc}\left[\mathbbm{1} + \Pi_\mathrm{imp}W_\mathrm{loc} \right]^{-1}.
\end{align}
From the solution of the impurity problem new $\Sigma^{\textrm{EDMFT}}$ and $\Pi^{\textrm{EDMFT}}$ are obtained, and the scheme is iterated until converge ($G_\textrm{loc}=G_\textrm{imp}$ and $W_\textrm{loc}=W_\textrm{imp}$), providing a self-consistent feedback between the local and nonlocal correlations.

We further use the approach of Ref.~\cite{Petocchi2020b} and split up the local correlated space (solved within EDMFT) into two separate impurity problems for each site in the unit cell.
The local quantities thus acquire an additional index $i$ for the site, $\Sigma^{\textrm{EDMFT}}_{\textrm{imp,}i}$ and $\Pi^{\textrm{EDMFT}}_{\textrm{imp,}i}$, and the $GW$+EDMFT self-consistency condition is modified to $G_\textrm{loc}^i=G_\textrm{imp}^i$ and $W_\textrm{loc}^i=W_\textrm{imp}^i$ (i.e. the local projections of $G$ and $W$ should be equal to the corresponding impurity quantities, for both sites). Due to the strong $sp$-bonds in the binary semiconductors studied, this implies an approximation, i.e., we do not consider the short-range correlations at the impurity level and therefore do not calculate EDMFT vertex corrections associated with \emph{intersite} terms.

\subsection{Computational details}

For the studied elemental and binary semiconductors and insulators we use the experimental structures and lattice parameters. We only focus on the materials in the cubic family, and for CdS and GaN use the cubic modification despite the wurtzite structure being the more stable one at ambient conditions \footnote{The wurtzite structure has two formula units in the primitive cell and requires a doubled model space.}.
We start from a DFT calculation 
using the \textsc{FLEUR} full-potential all-electron code \cite{fleurcode}. For the DFT calculations we use the GGA functional \cite{Perdew1996} and a $20\times20\times20$ {\bf k}-grid.
We then define a low-energy model by projecting Wannier functions \cite{Marzari1997,Mostofi2008} of $s$ and $p$ character on each of the two atomic sites in the unit cell using the \textsc{WANNIER90} code \cite{Mostofi2008}. This results in an 8-orbital model space, and in Wannier functions that are localized on the atomic sites, with local hybridization functions that are diagonal in orbital space for each site, which is a requirement for the impurity solver used in EDMFT \cite{Werner2006,Werner2007}
\footnote{We solve two separate impurity problems for the two sites in the unit cell, as described above. Intersite correlations are treated at the $GW$ level and we keep all nonlocal components in the self-consistency loop.}.
Systems with significant local interorbital hybridizations would be computationally more demanding. To minimize the sign problem in the Monte Carlo solution of the impurity problem, a local rotation can be used \cite{Petocchi2021,Petocchi2026}, at the expense of a more complex interaction term.

We next downfold the full Hilbert space to the model by performing a constrained random-phase approximation (cRPA) \cite{Aryasetiawan2004} and a one-shot $GW$ calculation \cite{Hedin1965}, which effectively includes the higher-energy states in the effective ``bare" interaction $U_{ijkl}(\omega,{\bf q})$ and ``bare" propagator $G^{0}_{ij}(\omega,{\bf k})$ of the model space. For the downfolding we use the SPEX code \cite{Friedrich2010} and perform the calculations on a $8\times8\times8$ {\bf k}-grid (except for Si where we use a $10\times10\times10$ grid) and include unoccupied bands up to $\sim 80$-100 eV in the calculation of the polarization and self-energy. The self-consistent calculations ($GW$+EDMFT and multitier sc$GW$) in the low-energy space are performed using the implementation of Refs.~\cite{Nilsson2017,Petocchi2020b}. Here, we employ the same grid as in the SPEX calculation, but a nonzero temperature, which will lead to a further temperature broadening effect of the self-energy and polarization. We cut off the fermionic and bosonic Matsubara axis frequencies at 300 eV, where the relevant quantities are approaching their asymptotic values.
For each of the two atomic sites in the unit cell we consider an impurity problem consisting of the $s$ and $p$ orbitals that is solved using a continuous-time Monte Carlo solver \cite{Werner2006,Hafermann2013} capable of treating dynamically screened interactions $\mathcal{U}(\omega)$ \cite{Werner2010}.
From these impurity calculations we extract the self-consistently calculated local vertex corrections $\Sigma^{\textrm{EDMFT}}_\textrm{imp}$ and $\Pi^{\textrm{EDMFT}}_\textrm{imp}$.
For more details on our implementation of the $GW$+EDMFT scheme, we refer to Refs.~\cite{Nilsson2017,Petocchi2020b}. 
To compare our results to standard $GW$ calculations, and to explicitly verify that the corrections to the observables are not resulting from truncating the states included in the self-consistent low-energy space, we also perform similar multitier $GW$ calculations (without the local vertex corrections from EDMFT in Eqs.~\eqref{Eq:Sigma} and~\eqref{Eq:Pi}).

\section{Results} The calculations in the low-energy space are performed at inverse temperature $\beta=30$ eV$^{-1}$ ($T\sim390$ K) \footnote{A small technical issue with the $\beta=30$ eV$^{-1}$ calculations is that the $G_0W_0$ downfolding to the low-energy space in the multitier scheme is performed at $T=0$ \cite{Friedrich2010}. This inconsistency leads to a slight doping of the conduction band, which we fix by manually placing the chemical potential in the gap.}
 and we checked for several materials that a further temperature decrease to $\beta=40$ eV$^{-1}$ ($T\sim 290$~K) does not affect our results. We can thus directly compare them to the experimental data measured near room temperature ($T\sim$~300~K).

\begin{figure}[t!] 
\begin{centering}
\includegraphics[width=0.9\columnwidth]{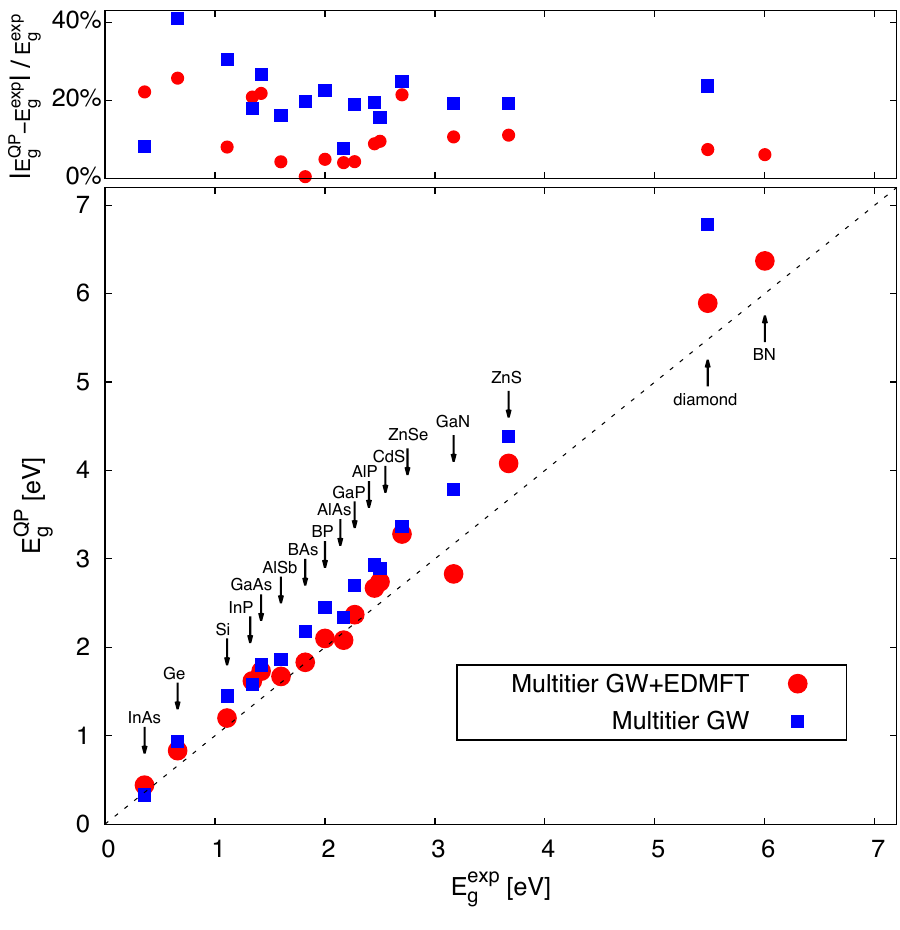}
\par\end{centering}
\caption{
Theoretically calculated quasiparticle band gaps (see text) versus experimentally measured band gaps for the indicated elemental and binary semiconductors and insulators. The theoretical values are for multitier $GW$+EDMFT (red circles) and $GW$ (blue squares). The $GW$ gap for BN is outside the shown range at 10 eV, see text.
Top panel: Relative error of the quasiparticle gaps compared to experiment.
The experimental gaps are taken from Refs.~\cite{Madelung1999,Madelung2002}.
 \label{Fig:QPgaps}
}
\end{figure}

The Matsubara-axis solution in the low-energy space does not give us direct access to the real-frequency self-energy and spectral function. We use the following two
independent approaches to extract the band gap:
\begin{itemize}
\item[(i)] a linearization of the self-energy around $\Sigma(i\omega_n~\to~0)$, 
\item[(ii)] Maximum Entropy (MaxEnt) analytical continuation \cite{Jarrell1996} of $G(i\omega_n)$ to obtain $A(\omega)=-\frac{1}{\pi}\text{Im}G(\omega)$.
\end{itemize}
Since there are different approximations involved in these two approaches, the comparison of the results provides a consistency check.

In procedure (i) we rewrite the quasiparticle equation \cite{Hybertsen1986}
by linearizing it around $\varepsilon_{n{\bf k}}$:
\begin{equation}
E^\textrm{QP}_{n{\bf k}} \approx \varepsilon_{n{\bf k}} + Z_{n{\bf k}}\big( \Sigma_{n{\bf k}}(0) - V_{n{\bf k}}^\mathrm{xc} \big),
\end{equation}
replacing $ \text{Re}[\Sigma_{n{\bf k}}(\varepsilon_{n{\bf k}}] )$ by $\Sigma_{n{\bf k}}(0)$, since the real part of the self-energy is rather flat in the energy range needed to estimate the band edges.
The renormalization factor $Z_{n{\bf k}}^{-1} = 1- \text{Im}\Sigma_{n{\bf k}}(i\omega_0)/\omega_0$,
is evaluated directly on the Matsubara axis, again using the linearity of $\text{Im}\Sigma_{n{\bf k}}$ in the relevant energy region. The gap is then directly calculated from the quasi-particle energy dispersion.
(Note that this approximation is merely used for the gap extraction, while in the self-consistent calculations we take into account all matrix elements and the full frequency-dependence of $\Sigma_{ij}(i\omega_n,{\bf k})$.)

In procedure (ii) we estimate the band gap from the local spectral function $A_\text{loc}(\omega)$, i.e., the energy range in which the spectral weight essentially vanishes.

In Fig.~\ref{Fig:QPgaps} we show the calculated band gaps for both the self-consistent multitier $GW$+EDMFT calculations, with the local vertex corrections included, and a multitier $GW$ calculation. As expected, the band gaps in the latter approach are overestimated, with errors as large as several eV for the large gap insulators (diamond and BN). 
We also note that our multitier $GW$ results, which only include a self-consistency within the $sp$ subspace, are in good agreement with sc$GW$ calculations that perform a self-consistency in the full Hilbert space \cite{Kutepov2012,Kutepov2017}.
The $GW$+EDMFT results show a remarkable improvement over the $GW$ results in Fig.~\ref{Fig:QPgaps}, with consistently smaller band gaps and excellent agreement with experiment for many systems. These results also compare favorably to those from other advanced methods, such as the well-established QSGW method, the vertex corrected QSGW scheme by Cunningham \emph{et al.} (QS$G\hat{W}$) \cite{Cunningham2023}, and the vertex corrected $GW$ by Kutepov \cite{Kutepov2017}, which are all considered well-suited for this class of weakly-correlated materials. An explicit comparison of the band gaps for a subset of compounds is shown in Table~\ref{Table:bandgaps}.

 \begin{table}[t]
\setlength{\tabcolsep}{1.8pt} 
\renewcommand{\arraystretch}{1.4} 
\caption{\label{Table:bandgaps} Band gaps in eV calculated using $GW$+EDMFT, compared to experiment and other related methods discussed in the text, for several materials considered in Fig.~\ref{Fig:QPgaps}. The QSGW and vc$GW$ values are taken from Ref.~\cite{Kutepov2017} and the (vertex corrected) QS$G\hat{W}$ values from Ref.~\cite{Cunningham2023} (we were unable to extract the gaps for AlP and BN).
}
 \begin{tabular}{|c||c c c c c|}
\hline
& QSGW & QS$G\hat{W}$ & vc$GW$  & $GW$+EDMFT & Exp \\ 
\hline\hline
Si &  1.41 & 1.08 & 1.26-1.32 & 1.20  &  1.17 \\
GaAs &  1.96 & 1.63 & 1.72-1.80 & 1.73  &  1.42 \\
AlP &  2.80 & & 2.44-2.53 & 2.67 & 2.45 \\ 
ZnS &  4.2 & 3.7 & 3.8-3.9 & 4.1 & 3.7 \\
C (diamond) &  6.2 & 5.6 & 5.7-5.8  & 5.9  & 5.5 \\
BN &  7.1 & & 6.3-6.4 & 6.4 & 6.0-6.2 \\ 
\hline
 \end{tabular}
 \end{table}

The reported quasiparticle gaps from method (i) are in overall good agreement with the MaxEnt spectral functions $A_\mathrm{loc}(\omega)$ (method (ii)), although it is more difficult to extract a precise gap value in the latter case due to the smearing of the band edges. 
Even for the wide gap insulators, where one might expect the largest error from the linearization of the self-energy around $\Sigma(i\omega\to0)$, we find a good agreement between the two estimates.
For example, the diamond spectral function (not shown) gives a gap $E^\textrm{MaxEnt}_g\approx$ $6.1\pm0.2$ eV,
which is in agreement with the linearized QP band gap ($E^\textrm{QP}=5.9$ eV), and close to the experimentally observed gap ($E^\textrm{exp}=5.5$ eV). The exception to this is multitier $GW$ for BN, where the MaxEnt spectrum gives an estimated gap of $7.7\pm0.2$ versus a quasiparticle gap of almost 10~eV.

\begin{figure}
\begin{centering}
\includegraphics[width=0.85\columnwidth]{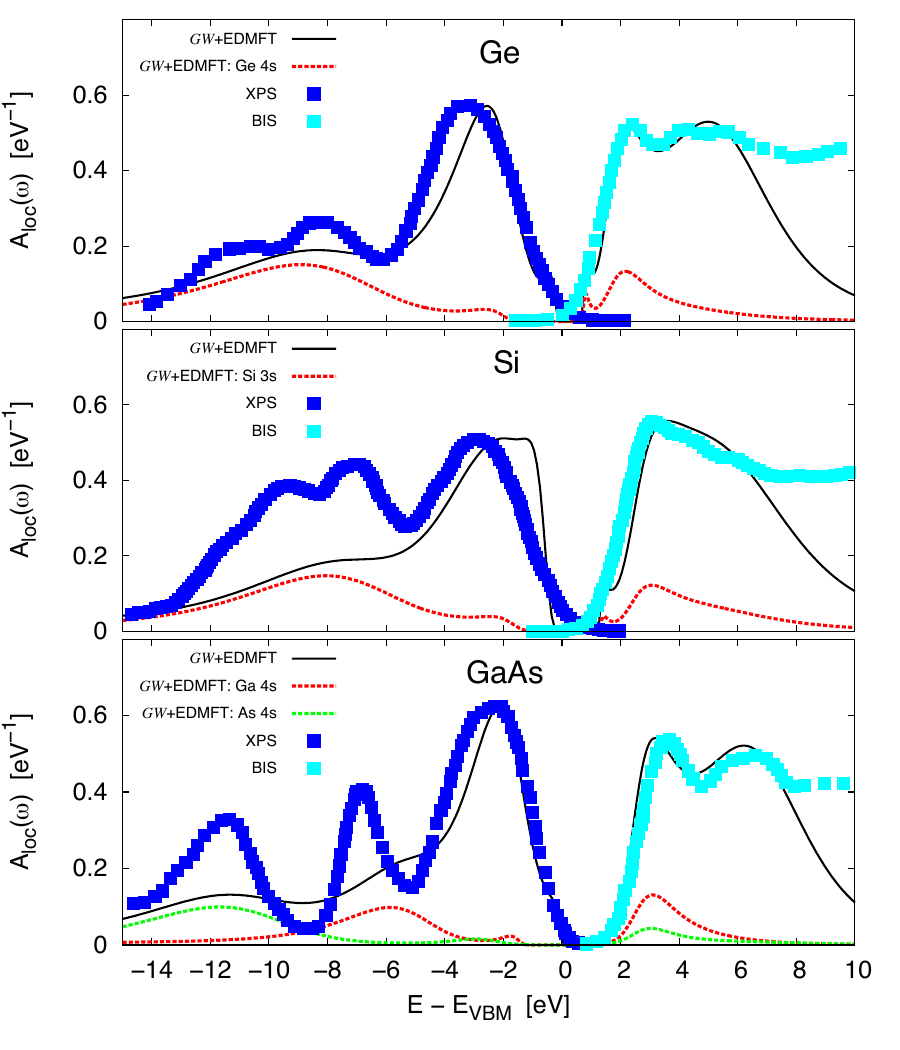}
\par\end{centering}
\caption{
Theoretically calculated spectral function $A_\text{loc}(\omega)$ from $GW$+EDMFT (black line) compared to experimental photoemission spectra (blue and cyan squares) from Ref.~\cite{Jackson1988}.
The theoretical chemical potential is set at the VBM, while the experimental data are shown as reported.
The contributions from the $s$ states are indicated by the dashed lines.
\label{Fig:Spectra}}
\end{figure}

The good prediction of the gap value and the conduction and valence band features
is further demonstrated in Fig.~\ref{Fig:Spectra}, where we compare our theoretical spectra to experimental results from x-ray photoemission spectroscopy (XPS) and bremsstrahlung isochromat spectroscopy (BIS) \cite{Jackson1988}.
Here we show the MaxEnt local spectral functions for heteropolar direct-gapped GaAs and homopolar indirect-gapped Ge and Si, to validate our method on prototypical semiconductors of different types.
We set the zero of energy at the valence band maximum (VBM) in the calculations, and compare the computed spectra to experimental data which have only been rescaled to match the peak heights in the valence and conduction bands.
The theoretically calculated and experimentally measured spectra agree very well in the peak positions, whereas the relative weights of the peaks differ.
This is not surprising:
we do not take into account selection rules or matrix elements of the dipole operator, which would be necessary for a quantitative comparison to experiment \cite{Hedin1999}.

Starting with Ge, the valence band and conduction band edges match well with the experimental data and the peak positions (quasiparticle energies) are in good agreement, although the dominant peak in the valence band is shifted somewhat up in energy and is slightly too narrow. The shoulder structure around $-11$ eV is missing or merged with the peak around $-8$ eV, possibly due to the limitations of MaxEnt. The agreement of the conduction band is markedly better, with the characteristic two-peak structure well reproduced up to 7-8 eV, which is the upper edge of our low-energy space. 
Similarly, in GaAs, the band edges and the two-peak structure in the conduction band reproduce the experimental spectrum well. In the valence band, the lowest and highest energy peaks (mostly $s$ and $p$ derived, respectively) are close to the experimental positions. The central peak, which originates from the $s$-$p$ hybridization, however, is around 1~eV too high in energy.
This may be explained by the use of atom centered Wannier functions and the fact that we treat short range intersite correlations at the $GW$ level only (the local EDMFT vertex corrections are taken from separate impurity problems for each site).
We find the worst agreement for Si, where although the conduction band is reproduced very well with a main peak and a broad shoulder, the theoretical valence band edge displays a sharp onset unlike the experimental data. As for Ge, the Si middle peak is merged with the lower peak.

Since the only difference between the $GW$ and $GW$+EDMFT calculations is the inclusion of local vertex corrections to the self-energy and polarization from EDMFT, our results  reveal the effect of such corrections on experimentally accessible quantities. In particular, they show that these corrections are essential for a quantitatively accurate description of the electronic structure, despite the weakly correlated nature of the materials. 

\begin{figure}[t]
\begin{centering}
\includegraphics[width=0.95\columnwidth]{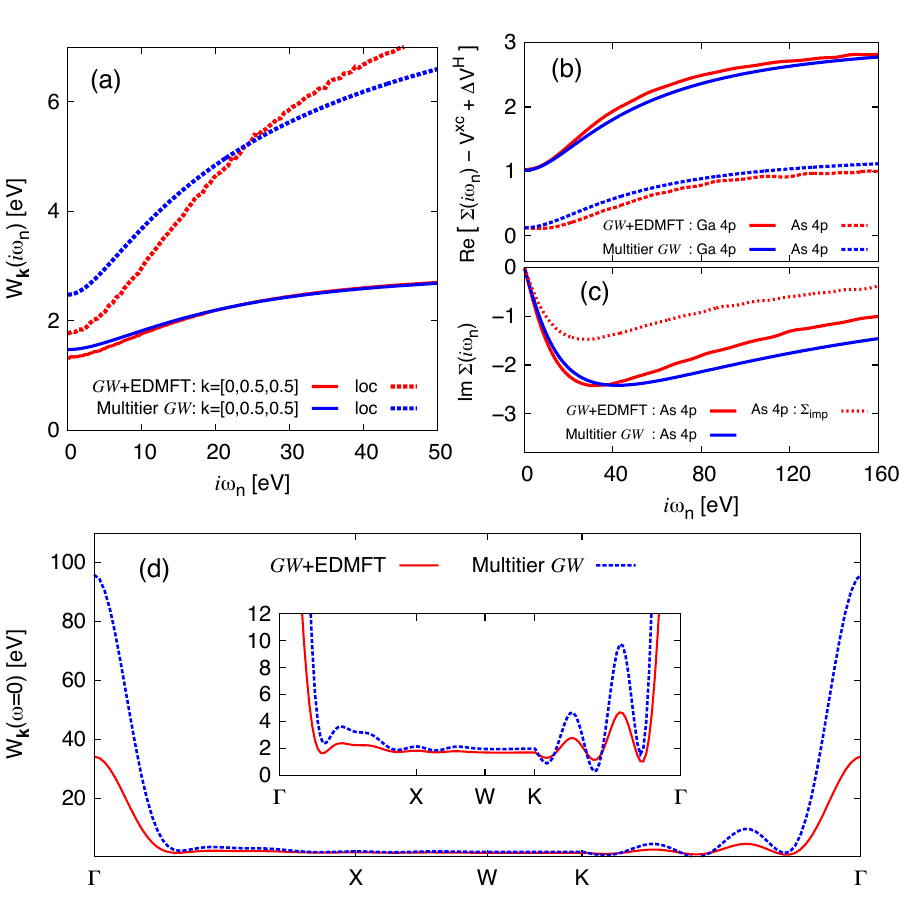}
\par\end{centering}
\caption{\label{Fig:Sigma_W}
(a) Frequency dependence of the screened interaction $W_\mathbf{k}(i\omega)$ at the indicated momentum transfer $\mathbf{k}$ and the local $W_\textrm{loc}$ for As $4p$.
Frequency-dependence of the real (b) and imaginary (c) parts of the local self-energy.
(d) The static $W_\mathbf{k}(\omega=0)$ along the indicated high-symmetry path.
}
\end{figure}

From the calculated QP energies $E_{n{\bf k}}^\textrm{QP}$, we can also extract the band masses, which yields more mixed results. As an example, the light and heavy hole masses in GaAs calculated
along the [100] direction are $0.088(5)m_0$ and $0.39(1)m_0$ in units of the electron mass $m_0$ in excellent agreement with the experimental values of 0.085-0.090$m_0$ and 0.33-0.45$m_0$ \cite{Madelung2002}. The electron effective mass of 0.085(5)$m_0$ is in a bit worse agreement with the experimental value ($0.064$-$0.069m_0$).
In the case of Ge, the masses of the light (calculated: $0.060(5)m_0$, exp: $0.044m_0$) and heavy (calculated: $0.25(2)m_0$, exp: $0.284m_0$) holes along the [100] direction show an acceptable agreement, whereas along [111] there is a large discrepancy for the heavy hole (calculated: $0.60(2)m_0$, exp: $0.376m_0$).
Also the electron effective mass at the L point along the transverse (calculated: $0.025(5)m_0$, exp: $0.082m_0$) and longitudinal (calculated: 1.1(1)$m_0$, exp: 1.54-1.74$m_0$) directions is not at all accurately reproduced.

Since this analysis is based on a linearized and interpolated self-energy (interpolated to nearby points), it is not too surprising that the band mass, which is sensitive to the details of the dispersion, is difficult to compute.
This is exacerbated by the fact that we perform our calculations at relatively high temperatures ($\sim$300 K), where the band dispersion clearly will become broadened, and the effective band masses therefore harder to calculate. In other words, it makes the interpolated self-energy at these interpolated points less reliable.
The errors in the calculated values quoted above, which are based on using different regions for the numerical derivative, may thus substantially underestimate the real errorbars, if the temperature broadening has a significant effect on the dispersion in the indicated directions. To clarify this issue, calculations would need to include these points explicilty in the {\bf k}-grids, or be performed at substantially lower temperatures, where these errors should be reduced.
These problems are absent in methods such as (vertex corrected) QS$G\hat{W}$ \cite{Cunningham2023}, which allow calculations at zero temperature, and which do find a substantially better agreement with the experimental masses.

The necessity of considering the dynamical screening for reliable calculations of semiconductors and insulators on the $GW$ level was realized early on, and it is well-established that this leads to a systematic decrease of the band gaps compared to a static approximation \cite{Hybertsen1986}.
The decrease of the calculated band gaps in $GW$+EDMFT, compared to the sc$GW$ results, can similarly be traced back to an increased screening at low energy. Both the local screened interaction $W_\textrm{loc}=\frac{1}{N_{\mathbf k}}\sum_{\mathbf k}W_\mathbf{k}$ and the momentum dependent $W_\mathbf{k}$ are reduced in $GW$+EDMFT, as demonstrated for GaAs in Fig.~\ref{Fig:Sigma_W}(a,d).
The latter reduction directly affects the nonlocal self-energy and hence the gap size.
The local component of the self-energy, on the other hand, is not much changed, since the
shift from the impurity self-energy is almost canceled out locally by the double counting terms and changes in the Hartree potential. We show this explicitly for GaAs in Fig.~\ref{Fig:Sigma_W}(b,c), where we plot $\frac{1}{N_{\bf k}}\sum_{\bf k}\left[ \Sigma_{\bf k}(i\omega_n) + \Delta V^{H}_{\bf k} - V^\textrm{xc}_{\bf k}\right]$  for the As $4p$ and Ga $4p$ orbitals, which dominate the valence and conduction band edges, respectively. This corresponds to the local projection of the interaction terms entering in the calculation of the Green's function,
$G_{\bf k}^{-1}(i\omega_n)=i\omega_n + \mu - \varepsilon_{\bf k} +V^\textrm{xc}_{\bf k} - \Sigma_{\bf k}(i\omega_n) - \Delta V^{H}_{\bf k},$
\noindent
(here $\Delta V^{H}$ refers to the change in the Hartree potential due to the reshuffling of charge in the self-consistency).

\begin{table}[t]
\setlength{\tabcolsep}{1.4pt} 
\renewcommand{\arraystretch}{1.4} 
\caption{\label{Table:3d}
Cation $3d$ semicore energy levels $E_{d}$, with respect to the VBM, obtained from the local spectral function $A_\text{loc}(\omega)$ for LDA, $G^0W^0$, and $GW$+EDMFT, compared to experiment and the vertex corrected QS$G\hat{W}$ and $GW\Gamma^1$ methods. The uncertainty in the $GW$+EDMFT results refers to different parameters used in the analytical continuation. $GW\Gamma^1$ refers to the first order vertex correction with a static $W$ from Ref.~\cite{Gruneis2014}. The (vertex corrected) QS$G\hat{W}$ values are taken from Ref.~\cite{Cunningham2023}.
}
\begin{tabular}{|c | c c c c c c |}
\hline

\hline
& LDA & $G^0W^0$ & QS$G\hat{W}$ & $GW\Gamma^1$ & $GW$+EDMFT & Exp. \\
\hline

\hline
\hline
ZnS & 6.3 & 7.2    & 7.7 &  $8.4\pm0.1$  & $9.8\pm0.2$    & 9.0$^{a,b}$ \\ 
ZnSe & 6.5 & 7.4   & 8.2 &  $8.6\pm0.1$  & $9.4\pm0.2$    & 9.2$^{a}$, 9.4$^{b}$ \\
InP & 14.1 & 15.4  & 15.9 & $16.9\pm0.1$ & $16.9 \pm 0.2$ & 16.8$^{a}$ \\ 
GaAs & 14.8 & 16.6 & 17.4 & $18.5\pm0.1$ & $19.2\pm0.2$   & 18.8$^{a,c}$ \\ 
\hline

\hline

\end{tabular}

\scriptsize{$a$: Ref.~\cite{Ley1974}, $b$: Ref.~\cite{Weidemann1992}, $c$:  Ref.~\cite{Kraut1983}}
\end{table}

Next we extend the self-consistent low-energy space to also include the cation semicore $d$ states \footnote{In this case we are limited to $\beta=10$ eV$^{-1}$ for computational reasons.} 
in ZnS, ZnSe, InP, and GaAs, with experimental values ranging between 9 and 19 eV below the valence band maximum (see Table~\ref{Table:3d}).
These $d$ states are very localized compared to the more extended $s$ and $p$ states, as is seen from the spreads of the Wannier functions with 2.7, 4.4, and 0.3~\AA$^2$ for the Zn $s$, $p$, and $d$ states in ZnSe, respectively. While it follows that the interaction strengths are considerably larger, the $d$ states are almost fully filled (e.g. 9.8 electrons in the Zn $d$), so that they should not have a large contribution to the correlation part of the self-energy. 
We thus only treat the $sp$ states in the two impurity problems, as in the previous calculations. The treatment of these shallow $d$ states on the $GW$ level is however known to be important due to the hybridization with the $sp$ orbitals \cite{Rinke2005}.

First, we note that in the $spd$ model the band gaps remain mostly unaffected compared to the $sp$-only calculations, where the contribution from the $d$ states were kept frozen at the $G^0W^0$ level. This implies that the main effect on the low-energy states 
is that the $d$ states are pushed down compared to the initial DFT calculation, which is achieved already at the one-shot level. 
It is however well-established that one-shot $G^0W^0$ still places the $d$ states too high in energy \cite{Aryasetiawan1996,Rohlfing1997,Miyake2006}, while various forms of self-consistent calculations push these states further down \cite{Rohlfing1997,Shishkin2007}.
We also observe this downward shift in $GW$+EDMFT, resulting in a significantly improved placement of the $d$ states, in excellent agreement with experiment (Table~\ref{Table:3d}).
Compared to the first order vertex correction scheme with static $W$ by Gr\"uneis \emph{et al.} \cite{Gruneis2014}, we obtain a similar, or even slightly more accurate placement of the $d$ states.
Our semicore energy level positions are substantially improved compared to the (vertex corrected) QS$G\hat{W}$ prediction.
These results, together with the band gaps in Fig.~\ref{Fig:QPgaps} and the spectral functions in Fig.~\ref{Fig:Spectra}, demonstrate that $GW$+EDMFT provides a remarkably good description of the electronic structure in a wide energy range for 
elemental and binary semiconductors and insulators.

\section{Conclusions}

The state-of-the-art methods typically used for semiconductors and band insulators, e.g. (vertex corrected) QSGW or vc$GW$ \cite{Schilfgaarde2006,Kutepov2017,Cunningham2023},
provide a reliable description of this class of materials and are in general numerically cheaper than the $GW$+EDMFT scheme employed here. 
If one has \textit{a priori} knowledge about the correlations in the targeted system, and the latter are weak, it can therefore be more practical to choose one of these well-established methods.
The conceptually appealing features of the multitier $GW$+EDMFT approach, however, are the implementation of a fully self-consistent calculation in the relevant low-energy space, and in-principle applicability of the scheme to materials with \textit{arbitrary} correlation strength. The second point has up to now not been clarified, since the method has been mainly applied to systems with strong correlations, where the local EDMFT contribution to the self-energy is dominant. In this study, we tested $GW$+EDMFT on a range of weakly correlated semiconductors and band insulators and found that the method works well for also this class of materials, demonstrating the reliability and wide applicability of this {\it ab-initio} scheme.

Among the calculated observables, especially the band gaps and semicore $d$ states are improved compared to self-consistent $GW$, as a result of the inclusion of local vertex corrections. The $GW$+EDMFT results agree well with experiment, and compare favorably with other contemporary many-body methods that are considered well-suited to tackle semiconductors and insulators. The extracted band masses are however not accurate when compared with experiment, which is likely an effect of the interpolation of the self-energy at finite temperature (with an associated temperature broadening) that is needed for their calculation.

Our investigation of semiconductors and band insulators showed that the local vertex corrections, provided here nonperturbatively to all orders by EDMFT, are relevant for accurate self-consistent many-body calculations beyond $GW$, to correct for the inadequate treatment of local correlations in the $GW$ approximation and RPA.
The corrections to the polarization function enhance the local and nonlocal screening, which results in the demonstrated improvements compared to standard sc$GW$ calculations. 

Missing ingredients in this powerful {\it ab initio} scheme are spin-orbit coupling and electron-phonon renormalization effects on the band gaps, as well as nonlocal vertex corrections. In future studies, it would also be interesting to quantify the effects of full charge self-consistency in the high-energy space.

\bibliography{main}%

\begin{thebibliography}{67}%
\makeatletter
\providecommand \@ifxundefined [1]{%
 \@ifx{#1\undefined}
}%
\providecommand \@ifnum [1]{%
 \ifnum #1\expandafter \@firstoftwo
 \else \expandafter \@secondoftwo
 \fi
}%
\providecommand \@ifx [1]{%
 \ifx #1\expandafter \@firstoftwo
 \else \expandafter \@secondoftwo
 \fi
}%
\providecommand \natexlab [1]{#1}%
\providecommand \enquote  [1]{``#1''}%
\providecommand \bibnamefont  [1]{#1}%
\providecommand \bibfnamefont [1]{#1}%
\providecommand \citenamefont [1]{#1}%
\providecommand \href@noop [0]{\@secondoftwo}%
\providecommand \href [0]{\begingroup \@sanitize@url \@href}%
\providecommand \@href[1]{\@@startlink{#1}\@@href}%
\providecommand \@@href[1]{\endgroup#1\@@endlink}%
\providecommand \@sanitize@url [0]{\catcode `\\12\catcode `\$12\catcode
  `\&12\catcode `\#12\catcode `\^12\catcode `\_12\catcode `\%12\relax}%
\providecommand \@@startlink[1]{}%
\providecommand \@@endlink[0]{}%
\providecommand \url  [0]{\begingroup\@sanitize@url \@url }%
\providecommand \@url [1]{\endgroup\@href {#1}{\urlprefix }}%
\providecommand \urlprefix  [0]{URL }%
\providecommand \Eprint [0]{\href }%
\providecommand \doibase [0]{https://doi.org/}%
\providecommand \selectlanguage [0]{\@gobble}%
\providecommand \bibinfo  [0]{\@secondoftwo}%
\providecommand \bibfield  [0]{\@secondoftwo}%
\providecommand \translation [1]{[#1]}%
\providecommand \BibitemOpen [0]{}%
\providecommand \bibitemStop [0]{}%
\providecommand \bibitemNoStop [0]{.\EOS\space}%
\providecommand \EOS [0]{\spacefactor3000\relax}%
\providecommand \BibitemShut  [1]{\csname bibitem#1\endcsname}%
\let\auto@bib@innerbib\@empty
\bibitem [{\citenamefont {Hohenberg}\ and\ \citenamefont
  {Kohn}(1964)}]{Hohenberg1964}%
  \BibitemOpen
  \bibfield  {author} {\bibinfo {author} {\bibfnamefont {P.}~\bibnamefont
  {Hohenberg}}\ and\ \bibinfo {author} {\bibfnamefont {W.}~\bibnamefont
  {Kohn}},\ }\bibfield  {title} {\bibinfo {title} {{Inhomogeneous Electron
  Gas}},\ }\href {https://doi.org/10.1103/PhysRev.136.B864} {\bibfield
  {journal} {\bibinfo  {journal} {Phys. Rev.}\ }\textbf {\bibinfo {volume}
  {136}},\ \bibinfo {pages} {B864} (\bibinfo {year} {1964})}\BibitemShut
  {NoStop}%
\bibitem [{\citenamefont {Kohn}\ and\ \citenamefont {Sham}(1965)}]{Kohn1965}%
  \BibitemOpen
  \bibfield  {author} {\bibinfo {author} {\bibfnamefont {W.}~\bibnamefont
  {Kohn}}\ and\ \bibinfo {author} {\bibfnamefont {L.~J.}\ \bibnamefont
  {Sham}},\ }\bibfield  {title} {\bibinfo {title} {{Self-Consistent Equations
  Including Exchange and Correlation Effects}},\ }\href
  {https://doi.org/10.1103/PhysRev.140.A1133} {\bibfield  {journal} {\bibinfo
  {journal} {Phys. Rev.}\ }\textbf {\bibinfo {volume} {140}},\ \bibinfo {pages}
  {A1133} (\bibinfo {year} {1965})}\BibitemShut {NoStop}%
\bibitem [{\citenamefont {Sham}\ and\ \citenamefont
  {Schl\"uter}(1983)}]{Sham1983}%
  \BibitemOpen
  \bibfield  {author} {\bibinfo {author} {\bibfnamefont {L.~J.}\ \bibnamefont
  {Sham}}\ and\ \bibinfo {author} {\bibfnamefont {M.}~\bibnamefont
  {Schl\"uter}},\ }\bibfield  {title} {\bibinfo {title} {{Density-Functional
  Theory of the Energy Gap}},\ }\href
  {https://doi.org/10.1103/PhysRevLett.51.1888} {\bibfield  {journal} {\bibinfo
   {journal} {Phys. Rev. Lett.}\ }\textbf {\bibinfo {volume} {51}},\ \bibinfo
  {pages} {1888} (\bibinfo {year} {1983})}\BibitemShut {NoStop}%
\bibitem [{\citenamefont {Xiao}\ \emph {et~al.}(2011)\citenamefont {Xiao},
  \citenamefont {Tahir-Kheli},\ and\ \citenamefont {Goddard~III}}]{Xiao2011}%
  \BibitemOpen
  \bibfield  {author} {\bibinfo {author} {\bibfnamefont {H.}~\bibnamefont
  {Xiao}}, \bibinfo {author} {\bibfnamefont {J.}~\bibnamefont {Tahir-Kheli}},\
  and\ \bibinfo {author} {\bibfnamefont {W.~A.}\ \bibnamefont {Goddard~III}},\
  }\bibfield  {title} {\bibinfo {title} {Accurate band gaps for semiconductors
  from density functional theory},\ }\href@noop {} {\bibfield  {journal}
  {\bibinfo  {journal} {The Journal of Physical Chemistry Letters}\ }\textbf
  {\bibinfo {volume} {2}},\ \bibinfo {pages} {212} (\bibinfo {year}
  {2011})}\BibitemShut {NoStop}%
\bibitem [{\citenamefont {Hedin}(1965)}]{Hedin1965}%
  \BibitemOpen
  \bibfield  {author} {\bibinfo {author} {\bibfnamefont {L.}~\bibnamefont
  {Hedin}},\ }\bibfield  {title} {\bibinfo {title} {{New Method for Calculating
  the One-Particle Green's Function with Application to the Electron-Gas
  Problem}},\ }\href {https://doi.org/10.1103/PhysRev.139.A796} {\bibfield
  {journal} {\bibinfo  {journal} {Phys. Rev.}\ }\textbf {\bibinfo {volume}
  {139}},\ \bibinfo {pages} {A796} (\bibinfo {year} {1965})}\BibitemShut
  {NoStop}%
\bibitem [{\citenamefont {Strinati}\ \emph {et~al.}(1980)\citenamefont
  {Strinati}, \citenamefont {Mattausch},\ and\ \citenamefont
  {Hanke}}]{Strinati1980}%
  \BibitemOpen
  \bibfield  {author} {\bibinfo {author} {\bibfnamefont {G.}~\bibnamefont
  {Strinati}}, \bibinfo {author} {\bibfnamefont {H.~J.}\ \bibnamefont
  {Mattausch}},\ and\ \bibinfo {author} {\bibfnamefont {W.}~\bibnamefont
  {Hanke}},\ }\bibfield  {title} {\bibinfo {title} {{Dynamical Correlation
  Effects on the Quasiparticle Bloch States of a Covalent Crystal}},\ }\href
  {https://doi.org/10.1103/PhysRevLett.45.290} {\bibfield  {journal} {\bibinfo
  {journal} {Phys. Rev. Lett.}\ }\textbf {\bibinfo {volume} {45}},\ \bibinfo
  {pages} {290} (\bibinfo {year} {1980})}\BibitemShut {NoStop}%
\bibitem [{\citenamefont {Strinati}\ \emph {et~al.}(1982)\citenamefont
  {Strinati}, \citenamefont {Mattausch},\ and\ \citenamefont
  {Hanke}}]{Strinati1982}%
  \BibitemOpen
  \bibfield  {author} {\bibinfo {author} {\bibfnamefont {G.}~\bibnamefont
  {Strinati}}, \bibinfo {author} {\bibfnamefont {H.~J.}\ \bibnamefont
  {Mattausch}},\ and\ \bibinfo {author} {\bibfnamefont {W.}~\bibnamefont
  {Hanke}},\ }\bibfield  {title} {\bibinfo {title} {{Dynamical aspects of
  correlation corrections in a covalent crystal}},\ }\href
  {https://doi.org/10.1103/PhysRevB.25.2867} {\bibfield  {journal} {\bibinfo
  {journal} {Phys. Rev. B}\ }\textbf {\bibinfo {volume} {25}},\ \bibinfo
  {pages} {2867} (\bibinfo {year} {1982})}\BibitemShut {NoStop}%
\bibitem [{\citenamefont {Hybertsen}\ and\ \citenamefont
  {Louie}(1986)}]{Hybertsen1986}%
  \BibitemOpen
  \bibfield  {author} {\bibinfo {author} {\bibfnamefont {M.~S.}\ \bibnamefont
  {Hybertsen}}\ and\ \bibinfo {author} {\bibfnamefont {S.~G.}\ \bibnamefont
  {Louie}},\ }\bibfield  {title} {\bibinfo {title} {{Electron correlation in
  semiconductors and insulators: Band gaps and quasiparticle energies}},\
  }\href {https://doi.org/10.1103/PhysRevB.34.5390} {\bibfield  {journal}
  {\bibinfo  {journal} {Phys. Rev. B}\ }\textbf {\bibinfo {volume} {34}},\
  \bibinfo {pages} {5390} (\bibinfo {year} {1986})}\BibitemShut {NoStop}%
\bibitem [{\citenamefont {van Schilfgaarde}\ \emph {et~al.}(2006)\citenamefont
  {van Schilfgaarde}, \citenamefont {Kotani},\ and\ \citenamefont
  {Faleev}}]{Schilfgaarde2006}%
  \BibitemOpen
  \bibfield  {author} {\bibinfo {author} {\bibfnamefont {M.}~\bibnamefont {van
  Schilfgaarde}}, \bibinfo {author} {\bibfnamefont {T.}~\bibnamefont
  {Kotani}},\ and\ \bibinfo {author} {\bibfnamefont {S.}~\bibnamefont
  {Faleev}},\ }\bibfield  {title} {\bibinfo {title} {{Quasiparticle
  Self-Consistent {$GW$} Theory}},\ }\href
  {https://doi.org/10.1103/PhysRevLett.96.226402} {\bibfield  {journal}
  {\bibinfo  {journal} {Phys. Rev. Lett.}\ }\textbf {\bibinfo {volume} {96}},\
  \bibinfo {pages} {226402} (\bibinfo {year} {2006})}\BibitemShut {NoStop}%
\bibitem [{\citenamefont {Kotani}\ \emph {et~al.}(2007)\citenamefont {Kotani},
  \citenamefont {van Schilfgaarde},\ and\ \citenamefont {Faleev}}]{Kotani2007}%
  \BibitemOpen
  \bibfield  {author} {\bibinfo {author} {\bibfnamefont {T.}~\bibnamefont
  {Kotani}}, \bibinfo {author} {\bibfnamefont {M.}~\bibnamefont {van
  Schilfgaarde}},\ and\ \bibinfo {author} {\bibfnamefont {S.~V.}\ \bibnamefont
  {Faleev}},\ }\bibfield  {title} {\bibinfo {title} {Quasiparticle
  self-consistent $gw$ method: A basis for the independent-particle
  approximation},\ }\href {https://doi.org/10.1103/PhysRevB.76.165106}
  {\bibfield  {journal} {\bibinfo  {journal} {Phys. Rev. B}\ }\textbf {\bibinfo
  {volume} {76}},\ \bibinfo {pages} {165106} (\bibinfo {year}
  {2007})}\BibitemShut {NoStop}%
\bibitem [{\citenamefont {Northrup}\ \emph {et~al.}(1987)\citenamefont
  {Northrup}, \citenamefont {Hybertsen},\ and\ \citenamefont
  {Louie}}]{Northrup1987}%
  \BibitemOpen
  \bibfield  {author} {\bibinfo {author} {\bibfnamefont {J.~E.}\ \bibnamefont
  {Northrup}}, \bibinfo {author} {\bibfnamefont {M.~S.}\ \bibnamefont
  {Hybertsen}},\ and\ \bibinfo {author} {\bibfnamefont {S.~G.}\ \bibnamefont
  {Louie}},\ }\bibfield  {title} {\bibinfo {title} {Theory of quasiparticle
  energies in alkali metals},\ }\href
  {https://doi.org/10.1103/PhysRevLett.59.819} {\bibfield  {journal} {\bibinfo
  {journal} {Phys. Rev. Lett.}\ }\textbf {\bibinfo {volume} {59}},\ \bibinfo
  {pages} {819} (\bibinfo {year} {1987})}\BibitemShut {NoStop}%
\bibitem [{\citenamefont {Luo}\ \emph {et~al.}(2002)\citenamefont {Luo},
  \citenamefont {Ismail-Beigi}, \citenamefont {Cohen},\ and\ \citenamefont
  {Louie}}]{Luo2002}%
  \BibitemOpen
  \bibfield  {author} {\bibinfo {author} {\bibfnamefont {W.}~\bibnamefont
  {Luo}}, \bibinfo {author} {\bibfnamefont {S.}~\bibnamefont {Ismail-Beigi}},
  \bibinfo {author} {\bibfnamefont {M.~L.}\ \bibnamefont {Cohen}},\ and\
  \bibinfo {author} {\bibfnamefont {S.~G.}\ \bibnamefont {Louie}},\ }\bibfield
  {title} {\bibinfo {title} {Quasiparticle band structure of zns and znse},\
  }\href {https://doi.org/10.1103/PhysRevB.66.195215} {\bibfield  {journal}
  {\bibinfo  {journal} {Phys. Rev. B}\ }\textbf {\bibinfo {volume} {66}},\
  \bibinfo {pages} {195215} (\bibinfo {year} {2002})}\BibitemShut {NoStop}%
\bibitem [{\citenamefont {Shishkin}\ and\ \citenamefont
  {Kresse}(2007)}]{Shishkin2007}%
  \BibitemOpen
  \bibfield  {author} {\bibinfo {author} {\bibfnamefont {M.}~\bibnamefont
  {Shishkin}}\ and\ \bibinfo {author} {\bibfnamefont {G.}~\bibnamefont
  {Kresse}},\ }\bibfield  {title} {\bibinfo {title} {Self-consistent {$GW$}
  calculations for semiconductors and insulators},\ }\href
  {https://doi.org/10.1103/PhysRevB.75.235102} {\bibfield  {journal} {\bibinfo
  {journal} {Phys. Rev. B}\ }\textbf {\bibinfo {volume} {75}},\ \bibinfo
  {pages} {235102} (\bibinfo {year} {2007})}\BibitemShut {NoStop}%
\bibitem [{\citenamefont {Stan}\ \emph {et~al.}(2009)\citenamefont {Stan},
  \citenamefont {Dahlen},\ and\ \citenamefont {van Leeuwen}}]{Stan2009}%
  \BibitemOpen
  \bibfield  {author} {\bibinfo {author} {\bibfnamefont {A.}~\bibnamefont
  {Stan}}, \bibinfo {author} {\bibfnamefont {N.~E.}\ \bibnamefont {Dahlen}},\
  and\ \bibinfo {author} {\bibfnamefont {R.}~\bibnamefont {van Leeuwen}},\
  }\bibfield  {title} {\bibinfo {title} {{Levels of self-consistency in the
  {$GW$} approximation}},\ }\href {https://doi.org/10.1063/1.3089567}
  {\bibfield  {journal} {\bibinfo  {journal} {The Journal of Chemical Physics}\
  }\textbf {\bibinfo {volume} {130}},\ \bibinfo {pages} {114105} (\bibinfo
  {year} {2009})}\BibitemShut {NoStop}%
\bibitem [{\citenamefont {Hellgren}\ \emph {et~al.}(2021)\citenamefont
  {Hellgren}, \citenamefont {Baguet}, \citenamefont {Calandra}, \citenamefont
  {Mauri},\ and\ \citenamefont {Wirtz}}]{Hellgren2021}%
  \BibitemOpen
  \bibfield  {author} {\bibinfo {author} {\bibfnamefont {M.}~\bibnamefont
  {Hellgren}}, \bibinfo {author} {\bibfnamefont {L.}~\bibnamefont {Baguet}},
  \bibinfo {author} {\bibfnamefont {M.}~\bibnamefont {Calandra}}, \bibinfo
  {author} {\bibfnamefont {F.}~\bibnamefont {Mauri}},\ and\ \bibinfo {author}
  {\bibfnamefont {L.}~\bibnamefont {Wirtz}},\ }\bibfield  {title} {\bibinfo
  {title} {Electronic structure of ${\mathrm{tise}}_{2}$ from a
  quasi-self-consistent ${G}_{0}{W}_{0}$ approach},\ }\href
  {https://doi.org/10.1103/PhysRevB.103.075101} {\bibfield  {journal} {\bibinfo
   {journal} {Phys. Rev. B}\ }\textbf {\bibinfo {volume} {103}},\ \bibinfo
  {pages} {075101} (\bibinfo {year} {2021})}\BibitemShut {NoStop}%
\bibitem [{\citenamefont {Gukelberger}\ \emph {et~al.}(2015)\citenamefont
  {Gukelberger}, \citenamefont {Huang},\ and\ \citenamefont
  {Werner}}]{Gukelberger2015}%
  \BibitemOpen
  \bibfield  {author} {\bibinfo {author} {\bibfnamefont {J.}~\bibnamefont
  {Gukelberger}}, \bibinfo {author} {\bibfnamefont {L.}~\bibnamefont {Huang}},\
  and\ \bibinfo {author} {\bibfnamefont {P.}~\bibnamefont {Werner}},\
  }\bibfield  {title} {\bibinfo {title} {{On the dangers of partial
  diagrammatic summations: Benchmarks for the two-dimensional Hubbard model in
  the weak-coupling regime}},\ }\href
  {https://doi.org/10.1103/PhysRevB.91.235114} {\bibfield  {journal} {\bibinfo
  {journal} {Phys. Rev. B}\ }\textbf {\bibinfo {volume} {91}},\ \bibinfo
  {pages} {235114} (\bibinfo {year} {2015})}\BibitemShut {NoStop}%
\bibitem [{\citenamefont {Kutepov}\ \emph {et~al.}(2012)\citenamefont
  {Kutepov}, \citenamefont {Haule}, \citenamefont {Savrasov},\ and\
  \citenamefont {Kotliar}}]{Kutepov2012}%
  \BibitemOpen
  \bibfield  {author} {\bibinfo {author} {\bibfnamefont {A.}~\bibnamefont
  {Kutepov}}, \bibinfo {author} {\bibfnamefont {K.}~\bibnamefont {Haule}},
  \bibinfo {author} {\bibfnamefont {S.~Y.}\ \bibnamefont {Savrasov}},\ and\
  \bibinfo {author} {\bibfnamefont {G.}~\bibnamefont {Kotliar}},\ }\bibfield
  {title} {\bibinfo {title} {{Electronic structure of Pu and Am metals by
  self-consistent relativistic {$GW$} method}},\ }\href
  {https://doi.org/10.1103/PhysRevB.85.155129} {\bibfield  {journal} {\bibinfo
  {journal} {Phys. Rev. B}\ }\textbf {\bibinfo {volume} {85}},\ \bibinfo
  {pages} {155129} (\bibinfo {year} {2012})}\BibitemShut {NoStop}%
\bibitem [{\citenamefont {Kutepov}(2017)}]{Kutepov2017}%
  \BibitemOpen
  \bibfield  {author} {\bibinfo {author} {\bibfnamefont {A.~L.}\ \bibnamefont
  {Kutepov}},\ }\bibfield  {title} {\bibinfo {title} {{Self-consistent solution
  of Hedin's equations: Semiconductors and insulators}},\ }\href
  {https://doi.org/10.1103/PhysRevB.95.195120} {\bibfield  {journal} {\bibinfo
  {journal} {Phys. Rev. B}\ }\textbf {\bibinfo {volume} {95}},\ \bibinfo
  {pages} {195120} (\bibinfo {year} {2017})}\BibitemShut {NoStop}%
\bibitem [{\citenamefont {Kutepov}(2016)}]{Kutepov2016}%
  \BibitemOpen
  \bibfield  {author} {\bibinfo {author} {\bibfnamefont {A.~L.}\ \bibnamefont
  {Kutepov}},\ }\bibfield  {title} {\bibinfo {title} {{Electronic structure of
  Na, K, Si, and LiF from self-consistent solution of Hedin's equations
  including vertex corrections}},\ }\href
  {https://doi.org/10.1103/PhysRevB.94.155101} {\bibfield  {journal} {\bibinfo
  {journal} {Phys. Rev. B}\ }\textbf {\bibinfo {volume} {94}},\ \bibinfo
  {pages} {155101} (\bibinfo {year} {2016})}\BibitemShut {NoStop}%
\bibitem [{\citenamefont {Nelson}\ \emph {et~al.}(2007)\citenamefont {Nelson},
  \citenamefont {Bokes}, \citenamefont {Rinke},\ and\ \citenamefont
  {Godby}}]{Nelson2007}%
  \BibitemOpen
  \bibfield  {author} {\bibinfo {author} {\bibfnamefont {W.}~\bibnamefont
  {Nelson}}, \bibinfo {author} {\bibfnamefont {P.}~\bibnamefont {Bokes}},
  \bibinfo {author} {\bibfnamefont {P.}~\bibnamefont {Rinke}},\ and\ \bibinfo
  {author} {\bibfnamefont {R.~W.}\ \bibnamefont {Godby}},\ }\bibfield  {title}
  {\bibinfo {title} {{Self-interaction in Green's-function theory of the
  hydrogen atom}},\ }\href {https://doi.org/10.1103/PhysRevA.75.032505}
  {\bibfield  {journal} {\bibinfo  {journal} {Phys. Rev. A}\ }\textbf {\bibinfo
  {volume} {75}},\ \bibinfo {pages} {032505} (\bibinfo {year}
  {2007})}\BibitemShut {NoStop}%
\bibitem [{\citenamefont {Aryasetiawan}\ \emph {et~al.}(2012)\citenamefont
  {Aryasetiawan}, \citenamefont {Sakuma},\ and\ \citenamefont
  {Karlsson}}]{Aryasetiawan2012}%
  \BibitemOpen
  \bibfield  {author} {\bibinfo {author} {\bibfnamefont {F.}~\bibnamefont
  {Aryasetiawan}}, \bibinfo {author} {\bibfnamefont {R.}~\bibnamefont
  {Sakuma}},\ and\ \bibinfo {author} {\bibfnamefont {K.}~\bibnamefont
  {Karlsson}},\ }\bibfield  {title} {\bibinfo {title} {{$GW$} approximation
  with self-screening correction},\ }\href
  {https://doi.org/10.1103/PhysRevB.85.035106} {\bibfield  {journal} {\bibinfo
  {journal} {Phys. Rev. B}\ }\textbf {\bibinfo {volume} {85}},\ \bibinfo
  {pages} {035106} (\bibinfo {year} {2012})}\BibitemShut {NoStop}%
\bibitem [{\citenamefont {Christiansson}\ and\ \citenamefont
  {Aryasetiawan}(2023)}]{Christiansson2023a}%
  \BibitemOpen
  \bibfield  {author} {\bibinfo {author} {\bibfnamefont {V.}~\bibnamefont
  {Christiansson}}\ and\ \bibinfo {author} {\bibfnamefont {F.}~\bibnamefont
  {Aryasetiawan}},\ }\bibfield  {title} {\bibinfo {title} {{Self-screening
  corrections beyond the random-phase approximation: Applications to band gaps
  of semiconductors}},\ }\href {https://doi.org/10.1103/PhysRevB.107.125105}
  {\bibfield  {journal} {\bibinfo  {journal} {Phys. Rev. B}\ }\textbf {\bibinfo
  {volume} {107}},\ \bibinfo {pages} {125105} (\bibinfo {year}
  {2023})}\BibitemShut {NoStop}%
\bibitem [{\citenamefont {Cunningham}\ \emph {et~al.}(2023)\citenamefont
  {Cunningham}, \citenamefont {Gr\"uning}, \citenamefont {Pashov},\ and\
  \citenamefont {van Schilfgaarde}}]{Cunningham2023}%
  \BibitemOpen
  \bibfield  {author} {\bibinfo {author} {\bibfnamefont {B.}~\bibnamefont
  {Cunningham}}, \bibinfo {author} {\bibfnamefont {M.}~\bibnamefont
  {Gr\"uning}}, \bibinfo {author} {\bibfnamefont {D.}~\bibnamefont {Pashov}},\
  and\ \bibinfo {author} {\bibfnamefont {M.}~\bibnamefont {van Schilfgaarde}},\
  }\bibfield  {title} {\bibinfo {title} {{{$\mathrm{QS}G\widehat{W}$}:
  Quasiparticle self-consistent {$GW$} with ladder diagrams in {$W$}}},\ }\href
  {https://doi.org/10.1103/PhysRevB.108.165104} {\bibfield  {journal} {\bibinfo
   {journal} {Phys. Rev. B}\ }\textbf {\bibinfo {volume} {108}},\ \bibinfo
  {pages} {165104} (\bibinfo {year} {2023})}\BibitemShut {NoStop}%
\bibitem [{\citenamefont {Gr\"uneis}\ \emph {et~al.}(2014)\citenamefont
  {Gr\"uneis}, \citenamefont {Kresse}, \citenamefont {Hinuma},\ and\
  \citenamefont {Oba}}]{Gruneis2014}%
  \BibitemOpen
  \bibfield  {author} {\bibinfo {author} {\bibfnamefont {A.}~\bibnamefont
  {Gr\"uneis}}, \bibinfo {author} {\bibfnamefont {G.}~\bibnamefont {Kresse}},
  \bibinfo {author} {\bibfnamefont {Y.}~\bibnamefont {Hinuma}},\ and\ \bibinfo
  {author} {\bibfnamefont {F.}~\bibnamefont {Oba}},\ }\bibfield  {title}
  {\bibinfo {title} {{Ionization Potentials of Solids: The Importance of Vertex
  Corrections}},\ }\href {https://doi.org/10.1103/PhysRevLett.112.096401}
  {\bibfield  {journal} {\bibinfo  {journal} {Phys. Rev. Lett.}\ }\textbf
  {\bibinfo {volume} {112}},\ \bibinfo {pages} {096401} (\bibinfo {year}
  {2014})}\BibitemShut {NoStop}%
\bibitem [{\citenamefont {Biermann}\ \emph {et~al.}(2003)\citenamefont
  {Biermann}, \citenamefont {Aryasetiawan},\ and\ \citenamefont
  {Georges}}]{Biermann2003}%
  \BibitemOpen
  \bibfield  {author} {\bibinfo {author} {\bibfnamefont {S.}~\bibnamefont
  {Biermann}}, \bibinfo {author} {\bibfnamefont {F.}~\bibnamefont
  {Aryasetiawan}},\ and\ \bibinfo {author} {\bibfnamefont {A.}~\bibnamefont
  {Georges}},\ }\bibfield  {title} {\bibinfo {title} {{First-Principles
  Approach to the Electronic Structure of Strongly Correlated Systems:
  Combining the {$GW$} Approximation and Dynamical Mean-Field Theory}},\ }\href
  {https://doi.org/10.1103/PhysRevLett.90.086402} {\bibfield  {journal}
  {\bibinfo  {journal} {Phys. Rev. Lett.}\ }\textbf {\bibinfo {volume} {90}},\
  \bibinfo {pages} {086402} (\bibinfo {year} {2003})}\BibitemShut {NoStop}%
\bibitem [{\citenamefont {Boehnke}\ \emph {et~al.}(2016)\citenamefont
  {Boehnke}, \citenamefont {Nilsson}, \citenamefont {Aryasetiawan},\ and\
  \citenamefont {Werner}}]{Boehnke2016}%
  \BibitemOpen
  \bibfield  {author} {\bibinfo {author} {\bibfnamefont {L.}~\bibnamefont
  {Boehnke}}, \bibinfo {author} {\bibfnamefont {F.}~\bibnamefont {Nilsson}},
  \bibinfo {author} {\bibfnamefont {F.}~\bibnamefont {Aryasetiawan}},\ and\
  \bibinfo {author} {\bibfnamefont {P.}~\bibnamefont {Werner}},\ }\bibfield
  {title} {\bibinfo {title} {{When strong correlations become weak: Consistent
  merging of {$GW$} and DMFT}},\ }\href
  {https://doi.org/10.1103/PhysRevB.94.201106} {\bibfield  {journal} {\bibinfo
  {journal} {Phys. Rev. B}\ }\textbf {\bibinfo {volume} {94}},\ \bibinfo
  {pages} {201106} (\bibinfo {year} {2016})}\BibitemShut {NoStop}%
\bibitem [{\citenamefont {Nilsson}\ \emph {et~al.}(2017)\citenamefont
  {Nilsson}, \citenamefont {Boehnke}, \citenamefont {Werner},\ and\
  \citenamefont {Aryasetiawan}}]{Nilsson2017}%
  \BibitemOpen
  \bibfield  {author} {\bibinfo {author} {\bibfnamefont {F.}~\bibnamefont
  {Nilsson}}, \bibinfo {author} {\bibfnamefont {L.}~\bibnamefont {Boehnke}},
  \bibinfo {author} {\bibfnamefont {P.}~\bibnamefont {Werner}},\ and\ \bibinfo
  {author} {\bibfnamefont {F.}~\bibnamefont {Aryasetiawan}},\ }\bibfield
  {title} {\bibinfo {title} {Multitier self-consistent {$GW+\text{EDMFT}$}},\
  }\href {https://doi.org/10.1103/PhysRevMaterials.1.043803} {\bibfield
  {journal} {\bibinfo  {journal} {Phys. Rev. Mater.}\ }\textbf {\bibinfo
  {volume} {1}},\ \bibinfo {pages} {043803} (\bibinfo {year}
  {2017})}\BibitemShut {NoStop}%
\bibitem [{\citenamefont {Petocchi}\ \emph
  {et~al.}(2020{\natexlab{a}})\citenamefont {Petocchi}, \citenamefont
  {Nilsson}, \citenamefont {Aryasetiawan},\ and\ \citenamefont
  {Werner}}]{Petocchi2020a}%
  \BibitemOpen
  \bibfield  {author} {\bibinfo {author} {\bibfnamefont {F.}~\bibnamefont
  {Petocchi}}, \bibinfo {author} {\bibfnamefont {F.}~\bibnamefont {Nilsson}},
  \bibinfo {author} {\bibfnamefont {F.}~\bibnamefont {Aryasetiawan}},\ and\
  \bibinfo {author} {\bibfnamefont {P.}~\bibnamefont {Werner}},\ }\bibfield
  {title} {\bibinfo {title} {{Screening from ${e}_{g}$ states and
  antiferromagnetic correlations in ${d}^{(1,2,3)}$ perovskites: A
  {$GW+\text{EDMFT}$} investigation}},\ }\href
  {https://doi.org/10.1103/PhysRevResearch.2.013191} {\bibfield  {journal}
  {\bibinfo  {journal} {Phys. Rev. Res.}\ }\textbf {\bibinfo {volume} {2}},\
  \bibinfo {pages} {013191} (\bibinfo {year} {2020}{\natexlab{a}})}\BibitemShut
  {NoStop}%
\bibitem [{\citenamefont {Kang}\ \emph {et~al.}()\citenamefont {Kang},
  \citenamefont {Semon}, \citenamefont {Melnick}, \citenamefont {Kotliar},\
  and\ \citenamefont {Choi}}]{Kang2024}%
  \BibitemOpen
  \bibfield  {author} {\bibinfo {author} {\bibfnamefont {B.}~\bibnamefont
  {Kang}}, \bibinfo {author} {\bibfnamefont {P.}~\bibnamefont {Semon}},
  \bibinfo {author} {\bibfnamefont {C.}~\bibnamefont {Melnick}}, \bibinfo
  {author} {\bibfnamefont {G.}~\bibnamefont {Kotliar}},\ and\ \bibinfo {author}
  {\bibfnamefont {S.}~\bibnamefont {Choi}},\ }\bibfield  {title} {\bibinfo
  {title} {{ComDMFT v.2.0: Fully Self-Consistent ab initio {$GW$+EDMFT} for the
  Electronic Structure of Correlated Quantum Materials}},\ }\href
  {https://arxiv.org/abs/2310.04613} {\bibinfo  {journal} {arxiv:2310.04613
  (2024)}\ }\BibitemShut {NoStop}%
\bibitem [{\citenamefont {Mushkaev}\ \emph {et~al.}(2024)\citenamefont
  {Mushkaev}, \citenamefont {Petocchi}, \citenamefont {Christiansson},\ and\
  \citenamefont {Werner}}]{Mushkaev2024}%
  \BibitemOpen
\bibfield  {journal} {  }\bibfield  {author} {\bibinfo {author} {\bibfnamefont
  {R.}~\bibnamefont {Mushkaev}}, \bibinfo {author} {\bibfnamefont
  {F.}~\bibnamefont {Petocchi}}, \bibinfo {author} {\bibfnamefont
  {V.}~\bibnamefont {Christiansson}},\ and\ \bibinfo {author} {\bibfnamefont
  {P.}~\bibnamefont {Werner}},\ }\bibfield  {title} {\bibinfo {title}
  {{Internal consistency of multi-tier {$GW$}+EDMFT}},\ }\href
  {https://doi.org/https://doi.org/10.1038/s41524-024-01376-6} {\bibfield
  {journal} {\bibinfo  {journal} {npj Comput Mater}\ }\textbf {\bibinfo
  {volume} {10}},\ \bibinfo {pages} {182} (\bibinfo {year} {2024})}\BibitemShut
  {NoStop}%
\bibitem [{\citenamefont {Petocchi}\ \emph {et~al.}(2021)\citenamefont
  {Petocchi}, \citenamefont {Christiansson},\ and\ \citenamefont
  {Werner}}]{Petocchi2021}%
  \BibitemOpen
  \bibfield  {author} {\bibinfo {author} {\bibfnamefont {F.}~\bibnamefont
  {Petocchi}}, \bibinfo {author} {\bibfnamefont {V.}~\bibnamefont
  {Christiansson}},\ and\ \bibinfo {author} {\bibfnamefont {P.}~\bibnamefont
  {Werner}},\ }\bibfield  {title} {\bibinfo {title} {Fully ab initio electronic
  structure of {${\mathrm{Ca}}_{2}{\mathrm{RuO}}_{4}$}},\ }\href
  {https://doi.org/10.1103/PhysRevB.104.195146} {\bibfield  {journal} {\bibinfo
   {journal} {Phys. Rev. B}\ }\textbf {\bibinfo {volume} {104}},\ \bibinfo
  {pages} {195146} (\bibinfo {year} {2021})}\BibitemShut {NoStop}%
\bibitem [{\citenamefont {Petocchi}\ \emph
  {et~al.}(2020{\natexlab{b}})\citenamefont {Petocchi}, \citenamefont
  {Christiansson}, \citenamefont {Nilsson}, \citenamefont {Aryasetiawan},\ and\
  \citenamefont {Werner}}]{Petocchi2020b}%
  \BibitemOpen
  \bibfield  {author} {\bibinfo {author} {\bibfnamefont {F.}~\bibnamefont
  {Petocchi}}, \bibinfo {author} {\bibfnamefont {V.}~\bibnamefont
  {Christiansson}}, \bibinfo {author} {\bibfnamefont {F.}~\bibnamefont
  {Nilsson}}, \bibinfo {author} {\bibfnamefont {F.}~\bibnamefont
  {Aryasetiawan}},\ and\ \bibinfo {author} {\bibfnamefont {P.}~\bibnamefont
  {Werner}},\ }\bibfield  {title} {\bibinfo {title} {{Normal State of
  {${\mathrm{Nd}}_{1\ensuremath{-}x}{\mathrm{Sr}}_{x}{\mathrm{NiO}}_{2}$} from
  Self-Consistent {$GW+\mathrm{EDMFT}$}}},\ }\href
  {https://doi.org/10.1103/PhysRevX.10.041047} {\bibfield  {journal} {\bibinfo
  {journal} {Phys. Rev. X}\ }\textbf {\bibinfo {volume} {10}},\ \bibinfo
  {pages} {041047} (\bibinfo {year} {2020}{\natexlab{b}})}\BibitemShut
  {NoStop}%
\bibitem [{\citenamefont {Christiansson}\ \emph
  {et~al.}(2023{\natexlab{a}})\citenamefont {Christiansson}, \citenamefont
  {Petocchi},\ and\ \citenamefont {Werner}}]{Christiansson2023b}%
  \BibitemOpen
  \bibfield  {author} {\bibinfo {author} {\bibfnamefont {V.}~\bibnamefont
  {Christiansson}}, \bibinfo {author} {\bibfnamefont {F.}~\bibnamefont
  {Petocchi}},\ and\ \bibinfo {author} {\bibfnamefont {P.}~\bibnamefont
  {Werner}},\ }\bibfield  {title} {\bibinfo {title} {{$GW+\mathrm{EDMFT}$}
  investigation of
  {${\mathrm{Pr}}_{1\ensuremath{-}x}{\mathrm{Sr}}_{x}{\mathrm{NiO}}_{2}$} under
  pressure},\ }\href {https://doi.org/10.1103/PhysRevB.107.045144} {\bibfield
  {journal} {\bibinfo  {journal} {Phys. Rev. B}\ }\textbf {\bibinfo {volume}
  {107}},\ \bibinfo {pages} {045144} (\bibinfo {year}
  {2023}{\natexlab{a}})}\BibitemShut {NoStop}%
\bibitem [{\citenamefont {Christiansson}\ \emph
  {et~al.}(2023{\natexlab{b}})\citenamefont {Christiansson}, \citenamefont
  {Petocchi},\ and\ \citenamefont {Werner}}]{Christiansson2023c}%
  \BibitemOpen
  \bibfield  {author} {\bibinfo {author} {\bibfnamefont {V.}~\bibnamefont
  {Christiansson}}, \bibinfo {author} {\bibfnamefont {F.}~\bibnamefont
  {Petocchi}},\ and\ \bibinfo {author} {\bibfnamefont {P.}~\bibnamefont
  {Werner}},\ }\bibfield  {title} {\bibinfo {title} {{Correlated Electronic
  Structure of ${\mathrm{La}}_{3}{\text{Ni}}_{2}{\mathrm{O}}_{7}$ under
  Pressure}},\ }\href {https://doi.org/10.1103/PhysRevLett.131.206501}
  {\bibfield  {journal} {\bibinfo  {journal} {Phys. Rev. Lett.}\ }\textbf
  {\bibinfo {volume} {131}},\ \bibinfo {pages} {206501} (\bibinfo {year}
  {2023}{\natexlab{b}})}\BibitemShut {NoStop}%
\bibitem [{\citenamefont {Christiansson}\ \emph {et~al.}(2022)\citenamefont
  {Christiansson}, \citenamefont {Petocchi},\ and\ \citenamefont
  {Werner}}]{Christiansson2022a}%
  \BibitemOpen
  \bibfield  {author} {\bibinfo {author} {\bibfnamefont {V.}~\bibnamefont
  {Christiansson}}, \bibinfo {author} {\bibfnamefont {F.}~\bibnamefont
  {Petocchi}},\ and\ \bibinfo {author} {\bibfnamefont {P.}~\bibnamefont
  {Werner}},\ }\bibfield  {title} {\bibinfo {title} {Superconductivity in black
  phosphorus and the role of dynamical screening},\ }\href
  {https://doi.org/10.1103/PhysRevB.105.174513} {\bibfield  {journal} {\bibinfo
   {journal} {Phys. Rev. B}\ }\textbf {\bibinfo {volume} {105}},\ \bibinfo
  {pages} {174513} (\bibinfo {year} {2022})}\BibitemShut {NoStop}%
\bibitem [{\citenamefont {Zhu}\ and\ \citenamefont {Chan}(2021)}]{Zhu2021}%
  \BibitemOpen
  \bibfield  {author} {\bibinfo {author} {\bibfnamefont {T.}~\bibnamefont
  {Zhu}}\ and\ \bibinfo {author} {\bibfnamefont {G.~K.-L.}\ \bibnamefont
  {Chan}},\ }\bibfield  {title} {\bibinfo {title} {{Ab Initio Full Cell
  {$GW+\mathrm{DMFT}$} for Correlated Materials}},\ }\href
  {https://doi.org/10.1103/PhysRevX.11.021006} {\bibfield  {journal} {\bibinfo
  {journal} {Phys. Rev. X}\ }\textbf {\bibinfo {volume} {11}},\ \bibinfo
  {pages} {021006} (\bibinfo {year} {2021})}\BibitemShut {NoStop}%
\bibitem [{\citenamefont {Almbladh}\ \emph {et~al.}(1999)\citenamefont
  {Almbladh}, \citenamefont {von Barth},\ and\ \citenamefont {van
  Leeuwen}}]{ALMBLADH1999}%
  \BibitemOpen
  \bibfield  {author} {\bibinfo {author} {\bibfnamefont {C.~O.}\ \bibnamefont
  {Almbladh}}, \bibinfo {author} {\bibfnamefont {U.}~\bibnamefont {von
  Barth}},\ and\ \bibinfo {author} {\bibfnamefont {R.}~\bibnamefont {van
  Leeuwen}},\ }\bibfield  {title} {\bibinfo {title} {Variational total energies
  from {$\Phi$}- and {$\Psi$}-derivable theories},\ }\href
  {https://doi.org/10.1142/S0217979299000436} {\bibfield  {journal} {\bibinfo
  {journal} {International Journal of Modern Physics B}\ }\textbf {\bibinfo
  {volume} {13}},\ \bibinfo {pages} {535} (\bibinfo {year} {1999})}\BibitemShut
  {NoStop}%
\bibitem [{\citenamefont {Chitra}\ and\ \citenamefont
  {Kotliar}(2001)}]{Chitra2001}%
  \BibitemOpen
  \bibfield  {author} {\bibinfo {author} {\bibfnamefont {R.}~\bibnamefont
  {Chitra}}\ and\ \bibinfo {author} {\bibfnamefont {G.}~\bibnamefont
  {Kotliar}},\ }\bibfield  {title} {\bibinfo {title} {{Effective-action
  approach to strongly correlated fermion systems}},\ }\href
  {https://doi.org/10.1103/PhysRevB.63.115110} {\bibfield  {journal} {\bibinfo
  {journal} {Phys. Rev. B}\ }\textbf {\bibinfo {volume} {63}},\ \bibinfo
  {pages} {115110} (\bibinfo {year} {2001})}\BibitemShut {NoStop}%
\bibitem [{\citenamefont {Georges}\ \emph {et~al.}(1996)\citenamefont
  {Georges}, \citenamefont {Kotliar}, \citenamefont {Krauth},\ and\
  \citenamefont {Rozenberg}}]{Georges1996}%
  \BibitemOpen
  \bibfield  {author} {\bibinfo {author} {\bibfnamefont {A.}~\bibnamefont
  {Georges}}, \bibinfo {author} {\bibfnamefont {G.}~\bibnamefont {Kotliar}},
  \bibinfo {author} {\bibfnamefont {W.}~\bibnamefont {Krauth}},\ and\ \bibinfo
  {author} {\bibfnamefont {M.~J.}\ \bibnamefont {Rozenberg}},\ }\bibfield
  {title} {\bibinfo {title} {Dynamical mean-field theory of strongly correlated
  fermion systems and the limit of infinite dimensions},\ }\href
  {https://doi.org/10.1103/RevModPhys.68.13} {\bibfield  {journal} {\bibinfo
  {journal} {Rev. Mod. Phys.}\ }\textbf {\bibinfo {volume} {68}},\ \bibinfo
  {pages} {13} (\bibinfo {year} {1996})}\BibitemShut {NoStop}%
\bibitem [{\citenamefont {Sun}\ and\ \citenamefont {Kotliar}(2002)}]{Sun2002}%
  \BibitemOpen
  \bibfield  {author} {\bibinfo {author} {\bibfnamefont {P.}~\bibnamefont
  {Sun}}\ and\ \bibinfo {author} {\bibfnamefont {G.}~\bibnamefont {Kotliar}},\
  }\bibfield  {title} {\bibinfo {title} {Extended dynamical mean-field theory
  and $\mathrm{GW}$ method},\ }\href
  {https://doi.org/10.1103/PhysRevB.66.085120} {\bibfield  {journal} {\bibinfo
  {journal} {Phys. Rev. B}\ }\textbf {\bibinfo {volume} {66}},\ \bibinfo
  {pages} {085120} (\bibinfo {year} {2002})}\BibitemShut {NoStop}%
\bibitem [{Note1()}]{Note1}%
  \BibitemOpen
  \bibinfo {note} {The wurtzite structure has two formula units in the
  primitive cell and requires a doubled model space.}\BibitemShut {Stop}%
\bibitem [{\citenamefont {{The FLEUR group}}()}]{fleurcode}%
  \BibitemOpen
  \bibfield  {author} {\bibinfo {author} {\bibnamefont {{The FLEUR group}}},\
  }\href@noop {} {\bibinfo {title} {{The FLEUR project}}},\ \bibinfo
  {howpublished} {\url{http://www.flapw.de}}\BibitemShut {NoStop}%
\bibitem [{\citenamefont {Perdew}\ \emph {et~al.}(1996)\citenamefont {Perdew},
  \citenamefont {Burke},\ and\ \citenamefont {Ernzerhof}}]{Perdew1996}%
  \BibitemOpen
  \bibfield  {author} {\bibinfo {author} {\bibfnamefont {J.~P.}\ \bibnamefont
  {Perdew}}, \bibinfo {author} {\bibfnamefont {K.}~\bibnamefont {Burke}},\ and\
  \bibinfo {author} {\bibfnamefont {M.}~\bibnamefont {Ernzerhof}},\ }\bibfield
  {title} {\bibinfo {title} {Generalized gradient approximation made simple},\
  }\href {https://doi.org/10.1103/PhysRevLett.77.3865} {\bibfield  {journal}
  {\bibinfo  {journal} {Phys. Rev. Lett.}\ }\textbf {\bibinfo {volume} {77}},\
  \bibinfo {pages} {3865} (\bibinfo {year} {1996})}\BibitemShut {NoStop}%
\bibitem [{\citenamefont {Marzari}\ and\ \citenamefont
  {Vanderbilt}(1997)}]{Marzari1997}%
  \BibitemOpen
  \bibfield  {author} {\bibinfo {author} {\bibfnamefont {N.}~\bibnamefont
  {Marzari}}\ and\ \bibinfo {author} {\bibfnamefont {D.}~\bibnamefont
  {Vanderbilt}},\ }\bibfield  {title} {\bibinfo {title} {{Maximally localized
  generalized Wannier functions for composite energy bands}},\ }\href
  {https://doi.org/10.1103/PhysRevB.56.12847} {\bibfield  {journal} {\bibinfo
  {journal} {{Phys. Rev. B}}\ }\textbf {\bibinfo {volume} {56}},\ \bibinfo
  {pages} {12847} (\bibinfo {year} {1997})}\BibitemShut {NoStop}%
\bibitem [{\citenamefont {Mostofi}\ \emph {et~al.}(2008)\citenamefont
  {Mostofi}, \citenamefont {Yates}, \citenamefont {Lee}, \citenamefont {Souza},
  \citenamefont {Vanderbilt},\ and\ \citenamefont {Marzari}}]{Mostofi2008}%
  \BibitemOpen
  \bibfield  {author} {\bibinfo {author} {\bibfnamefont {A.~A.}\ \bibnamefont
  {Mostofi}}, \bibinfo {author} {\bibfnamefont {J.~R.}\ \bibnamefont {Yates}},
  \bibinfo {author} {\bibfnamefont {Y.-S.}\ \bibnamefont {Lee}}, \bibinfo
  {author} {\bibfnamefont {I.}~\bibnamefont {Souza}}, \bibinfo {author}
  {\bibfnamefont {D.}~\bibnamefont {Vanderbilt}},\ and\ \bibinfo {author}
  {\bibfnamefont {N.}~\bibnamefont {Marzari}},\ }\bibfield  {title} {\bibinfo
  {title} {{wannier90: A tool for obtaining maximally-localised Wannier
  functions }},\ }\href
  {https://doi.org/{http://dx.doi.org/10.1016/j.cpc.2007.11.016}} {\bibfield
  {journal} {\bibinfo  {journal} {{Computer Physics Communications }}\ }\textbf
  {\bibinfo {volume} {178}},\ \bibinfo {pages} {685} (\bibinfo {year}
  {2008})}\BibitemShut {NoStop}%
\bibitem [{\citenamefont {Werner}\ \emph {et~al.}(2006)\citenamefont {Werner},
  \citenamefont {Comanac}, \citenamefont {de' Medici}, \citenamefont {Troyer},\
  and\ \citenamefont {Millis}}]{Werner2006}%
  \BibitemOpen
  \bibfield  {author} {\bibinfo {author} {\bibfnamefont {P.}~\bibnamefont
  {Werner}}, \bibinfo {author} {\bibfnamefont {A.}~\bibnamefont {Comanac}},
  \bibinfo {author} {\bibfnamefont {L.}~\bibnamefont {de' Medici}}, \bibinfo
  {author} {\bibfnamefont {M.}~\bibnamefont {Troyer}},\ and\ \bibinfo {author}
  {\bibfnamefont {A.~J.}\ \bibnamefont {Millis}},\ }\bibfield  {title}
  {\bibinfo {title} {{Continuous-Time Solver for Quantum Impurity Models}},\
  }\href {https://doi.org/10.1103/PhysRevLett.97.076405} {\bibfield  {journal}
  {\bibinfo  {journal} {Phys. Rev. Lett.}\ }\textbf {\bibinfo {volume} {97}},\
  \bibinfo {pages} {076405} (\bibinfo {year} {2006})}\BibitemShut {NoStop}%
\bibitem [{\citenamefont {Werner}\ and\ \citenamefont
  {Millis}(2007)}]{Werner2007}%
  \BibitemOpen
  \bibfield  {author} {\bibinfo {author} {\bibfnamefont {P.}~\bibnamefont
  {Werner}}\ and\ \bibinfo {author} {\bibfnamefont {A.~J.}\ \bibnamefont
  {Millis}},\ }\bibfield  {title} {\bibinfo {title} {{Efficient Dynamical Mean
  Field Simulation of the Holstein-Hubbard Model}},\ }\href
  {https://doi.org/10.1103/PhysRevLett.99.146404} {\bibfield  {journal}
  {\bibinfo  {journal} {Phys. Rev. Lett.}\ }\textbf {\bibinfo {volume} {99}},\
  \bibinfo {pages} {146404} (\bibinfo {year} {2007})}\BibitemShut {NoStop}%
\bibitem [{Note2()}]{Note2}%
  \BibitemOpen
  \bibinfo {note} {We solve two separate impurity problems for the two sites in
  the unit cell, as described above. Intersite correlations are treated at the
  $GW$ level and we keep all nonlocal components in the self-consistency
  loop.}\BibitemShut {Stop}%
\bibitem [{\citenamefont {Petocchi}\ \emph {et~al.}()\citenamefont {Petocchi},
  \citenamefont {Christiansson}, \citenamefont {Arita},\ and\ \citenamefont
  {Werner}}]{Petocchi2026}%
  \BibitemOpen
  \bibfield  {author} {\bibinfo {author} {\bibfnamefont {F.}~\bibnamefont
  {Petocchi}}, \bibinfo {author} {\bibfnamefont {V.}~\bibnamefont
  {Christiansson}}, \bibinfo {author} {\bibfnamefont {R.}~\bibnamefont
  {Arita}},\ and\ \bibinfo {author} {\bibfnamefont {P.}~\bibnamefont
  {Werner}},\ }\href {https://arxiv.org/abs/2601.18472} {\bibinfo {title}
  {{Plasmon assisted superconductivity in LiTi$_2$O$_4$}}},\ \Eprint
  {https://arxiv.org/abs/2601.18472 (2026)} {arXiv:2601.18472 (2026)}
  \BibitemShut {NoStop}%
\bibitem [{\citenamefont {Aryasetiawan}\ \emph {et~al.}(2004)\citenamefont
  {Aryasetiawan}, \citenamefont {Imada}, \citenamefont {Georges}, \citenamefont
  {Kotliar}, \citenamefont {Biermann},\ and\ \citenamefont
  {Lichtenstein}}]{Aryasetiawan2004}%
  \BibitemOpen
  \bibfield  {author} {\bibinfo {author} {\bibfnamefont {F.}~\bibnamefont
  {Aryasetiawan}}, \bibinfo {author} {\bibfnamefont {M.}~\bibnamefont {Imada}},
  \bibinfo {author} {\bibfnamefont {A.}~\bibnamefont {Georges}}, \bibinfo
  {author} {\bibfnamefont {G.}~\bibnamefont {Kotliar}}, \bibinfo {author}
  {\bibfnamefont {S.}~\bibnamefont {Biermann}},\ and\ \bibinfo {author}
  {\bibfnamefont {A.~I.}\ \bibnamefont {Lichtenstein}},\ }\bibfield  {title}
  {\bibinfo {title} {{Frequency-dependent local interactions and low-energy
  effective models from electronic structure calculations}},\ }\href
  {https://doi.org/10.1103/PhysRevB.70.195104} {\bibfield  {journal} {\bibinfo
  {journal} {Phys. Rev. B}\ }\textbf {\bibinfo {volume} {70}},\ \bibinfo
  {pages} {195104} (\bibinfo {year} {2004})}\BibitemShut {NoStop}%
\bibitem [{\citenamefont {Friedrich}\ \emph {et~al.}(2010)\citenamefont
  {Friedrich}, \citenamefont {Bl{\"{u}}gel},\ and\ \citenamefont
  {Schindlmayr}}]{Friedrich2010}%
  \BibitemOpen
  \bibfield  {author} {\bibinfo {author} {\bibfnamefont {C.}~\bibnamefont
  {Friedrich}}, \bibinfo {author} {\bibfnamefont {S.}~\bibnamefont
  {Bl{\"{u}}gel}},\ and\ \bibinfo {author} {\bibfnamefont {A.}~\bibnamefont
  {Schindlmayr}},\ }\bibfield  {title} {\bibinfo {title} {{Efficient
  implementation of the {$GW$} approximation within the all-electron FLAPW
  method}},\ }\href {https://doi.org/10.1103/PhysRevB.81.125102} {\bibfield
  {journal} {\bibinfo  {journal} {{Phys. Rev. B}}\ }\textbf {\bibinfo {volume}
  {81}},\ \bibinfo {pages} {125102} (\bibinfo {year} {2010})}\BibitemShut
  {NoStop}%
\bibitem [{\citenamefont {Hafermann}\ \emph {et~al.}(2013)\citenamefont
  {Hafermann}, \citenamefont {Werner},\ and\ \citenamefont
  {Gull}}]{Hafermann2013}%
  \BibitemOpen
  \bibfield  {author} {\bibinfo {author} {\bibfnamefont {H.}~\bibnamefont
  {Hafermann}}, \bibinfo {author} {\bibfnamefont {P.}~\bibnamefont {Werner}},\
  and\ \bibinfo {author} {\bibfnamefont {E.}~\bibnamefont {Gull}},\ }\bibfield
  {title} {\bibinfo {title} {Efficient implementation of the continuous-time
  hybridization expansion quantum impurity solver},\ }\href
  {https://doi.org/http://dx.doi.org/10.1016/j.cpc.2012.12.013} {\bibfield
  {journal} {\bibinfo  {journal} {Comput. Phys. Commun.}\ }\textbf {\bibinfo
  {volume} {184}},\ \bibinfo {pages} {1280 } (\bibinfo {year}
  {2013})}\BibitemShut {NoStop}%
\bibitem [{\citenamefont {Werner}\ and\ \citenamefont
  {Millis}(2010)}]{Werner2010}%
  \BibitemOpen
  \bibfield  {author} {\bibinfo {author} {\bibfnamefont {P.}~\bibnamefont
  {Werner}}\ and\ \bibinfo {author} {\bibfnamefont {A.~J.}\ \bibnamefont
  {Millis}},\ }\bibfield  {title} {\bibinfo {title} {Dynamical screening in
  correlated electron materials},\ }\href
  {https://doi.org/10.1103/PhysRevLett.104.146401} {\bibfield  {journal}
  {\bibinfo  {journal} {Phys. Rev. Lett.}\ }\textbf {\bibinfo {volume} {104}},\
  \bibinfo {pages} {146401} (\bibinfo {year} {2010})}\BibitemShut {NoStop}%
\bibitem [{Note3()}]{Note3}%
  \BibitemOpen
  \bibinfo {note} {A small technical issue with the $\beta =30$ eV$^{-1}$
  calculations is that the $G_0W_0$ downfolding to the low-energy space in the
  multitier scheme is performed at $T=0$ \cite {Friedrich2010}. This
  inconsistency leads to a slight doping of the conduction band, which we fix
  by manually placing the chemical potential in the gap.}\BibitemShut {Stop}%
\bibitem [{\citenamefont {Madelung}\ \emph {et~al.}(1999)\citenamefont
  {Madelung}, \citenamefont {R{\"o}ssler},\ and\ \citenamefont
  {Schulz}}]{Madelung1999}%
  \BibitemOpen
  \bibinfo {editor} {\bibfnamefont {O.}~\bibnamefont {Madelung}}, \bibinfo
  {editor} {\bibfnamefont {U.}~\bibnamefont {R{\"o}ssler}},\ and\ \bibinfo
  {editor} {\bibfnamefont {M.}~\bibnamefont {Schulz}},\ eds.,\ \href@noop {}
  {\emph {\bibinfo {title} {{II-VI and I-VII Compounds; Semimagnetic
  Compounds}}}},\ \bibinfo {series} {Landolt-B{\"o}rnstein - Group III
  Condensed Matter}, Vol.\ \bibinfo {volume} {41B}\ (\bibinfo  {publisher}
  {Springer-Verlag},\ \bibinfo {address} {Berlin Heidelberg},\ \bibinfo {year}
  {1999})\BibitemShut {NoStop}%
\bibitem [{\citenamefont {Madelung}\ \emph {et~al.}(2002)\citenamefont
  {Madelung}, \citenamefont {R{\"o}ssler},\ and\ \citenamefont
  {Schulz}}]{Madelung2002}%
  \BibitemOpen
  \bibinfo {editor} {\bibfnamefont {O.}~\bibnamefont {Madelung}}, \bibinfo
  {editor} {\bibfnamefont {U.}~\bibnamefont {R{\"o}ssler}},\ and\ \bibinfo
  {editor} {\bibfnamefont {M.}~\bibnamefont {Schulz}},\ eds.,\ \href@noop {}
  {\emph {\bibinfo {title} {{Group IV Elements, IV-IV and III-V Compounds. Part
  b - Electronic, Transport, Optical and Other Properties}}}},\ \bibinfo
  {series} {Landolt-B{\"o}rnstein - Group III Condensed Matter}, Vol.\ \bibinfo
  {volume} {41A1$\beta$}\ (\bibinfo  {publisher} {Springer-Verlag},\ \bibinfo
  {address} {Berlin Heidelberg},\ \bibinfo {year} {2002})\BibitemShut {NoStop}%
\bibitem [{\citenamefont {Jarrell}\ and\ \citenamefont
  {Gubernatis}(1996)}]{Jarrell1996}%
  \BibitemOpen
  \bibfield  {author} {\bibinfo {author} {\bibfnamefont {M.}~\bibnamefont
  {Jarrell}}\ and\ \bibinfo {author} {\bibfnamefont {J.}~\bibnamefont
  {Gubernatis}},\ }\bibfield  {title} {\bibinfo {title} {{Bayesian inference
  and the analytic continuation of imaginary-time quantum Monte Carlo data}},\
  }\href {https://doi.org/https://doi.org/10.1016/0370-1573(95)00074-7}
  {\bibfield  {journal} {\bibinfo  {journal} {Physics Reports}\ }\textbf
  {\bibinfo {volume} {269}},\ \bibinfo {pages} {133} (\bibinfo {year}
  {1996})}\BibitemShut {NoStop}%
\bibitem [{\citenamefont {Jackson}\ and\ \citenamefont
  {Allen}(1988)}]{Jackson1988}%
  \BibitemOpen
  \bibfield  {author} {\bibinfo {author} {\bibfnamefont {W.~B.}\ \bibnamefont
  {Jackson}}\ and\ \bibinfo {author} {\bibfnamefont {J.~W.}\ \bibnamefont
  {Allen}},\ }\bibfield  {title} {\bibinfo {title} {Experimental self-energy
  corrections for various semiconductors determined by electron spectroscopy},\
  }\href {https://doi.org/10.1103/PhysRevB.37.4618} {\bibfield  {journal}
  {\bibinfo  {journal} {Phys. Rev. B}\ }\textbf {\bibinfo {volume} {37}},\
  \bibinfo {pages} {4618} (\bibinfo {year} {1988})}\BibitemShut {NoStop}%
\bibitem [{\citenamefont {Hedin}(1999)}]{Hedin1999}%
  \BibitemOpen
  \bibfield  {author} {\bibinfo {author} {\bibfnamefont {L.}~\bibnamefont
  {Hedin}},\ }\bibfield  {title} {\bibinfo {title} {On correlation effects in
  electron spectroscopies and the {$GW$} approximation},\ }\href
  {https://doi.org/10.1088/0953-8984/11/42/201} {\bibfield  {journal} {\bibinfo
   {journal} {Journal of Physics: Condensed Matter}\ }\textbf {\bibinfo
  {volume} {11}},\ \bibinfo {pages} {R489} (\bibinfo {year}
  {1999})}\BibitemShut {NoStop}%
\bibitem [{\citenamefont {Ley}\ \emph {et~al.}(1974)\citenamefont {Ley},
  \citenamefont {Pollak}, \citenamefont {McFeely}, \citenamefont {Kowalczyk},\
  and\ \citenamefont {Shirley}}]{Ley1974}%
  \BibitemOpen
  \bibfield  {author} {\bibinfo {author} {\bibfnamefont {L.}~\bibnamefont
  {Ley}}, \bibinfo {author} {\bibfnamefont {R.~A.}\ \bibnamefont {Pollak}},
  \bibinfo {author} {\bibfnamefont {F.~R.}\ \bibnamefont {McFeely}}, \bibinfo
  {author} {\bibfnamefont {S.~P.}\ \bibnamefont {Kowalczyk}},\ and\ \bibinfo
  {author} {\bibfnamefont {D.~A.}\ \bibnamefont {Shirley}},\ }\bibfield
  {title} {\bibinfo {title} {{Total valence-band densities of states of III-V
  and II-VI compounds from x-ray photoemission spectroscopy}},\ }\href
  {https://doi.org/10.1103/PhysRevB.9.600} {\bibfield  {journal} {\bibinfo
  {journal} {Phys. Rev. B}\ }\textbf {\bibinfo {volume} {9}},\ \bibinfo {pages}
  {600} (\bibinfo {year} {1974})}\BibitemShut {NoStop}%
\bibitem [{\citenamefont {Weidemann}\ \emph {et~al.}(1992)\citenamefont
  {Weidemann}, \citenamefont {Gumlich}, \citenamefont {Kupsch}, \citenamefont
  {Middelmann},\ and\ \citenamefont {Becker}}]{Weidemann1992}%
  \BibitemOpen
  \bibfield  {author} {\bibinfo {author} {\bibfnamefont {R.}~\bibnamefont
  {Weidemann}}, \bibinfo {author} {\bibfnamefont {H.-E.}\ \bibnamefont
  {Gumlich}}, \bibinfo {author} {\bibfnamefont {M.}~\bibnamefont {Kupsch}},
  \bibinfo {author} {\bibfnamefont {H.-U.}\ \bibnamefont {Middelmann}},\ and\
  \bibinfo {author} {\bibfnamefont {U.}~\bibnamefont {Becker}},\ }\bibfield
  {title} {\bibinfo {title} {{Partial density of Mn 3d states and
  exchange-splitting changes in
  ${\mathrm{Zn}}_{1\mathrm{\ensuremath{-}}\mathit{x}}$${\mathrm{Mn}}_{\mathit{x}}$Y
  (Y=S,Se,Te)}},\ }\href {https://doi.org/10.1103/PhysRevB.45.1172} {\bibfield
  {journal} {\bibinfo  {journal} {Phys. Rev. B}\ }\textbf {\bibinfo {volume}
  {45}},\ \bibinfo {pages} {1172} (\bibinfo {year} {1992})}\BibitemShut
  {NoStop}%
\bibitem [{\citenamefont {Kraut}\ \emph {et~al.}(1983)\citenamefont {Kraut},
  \citenamefont {Grant}, \citenamefont {Waldrop},\ and\ \citenamefont
  {Kowalczyk}}]{Kraut1983}%
  \BibitemOpen
  \bibfield  {author} {\bibinfo {author} {\bibfnamefont {E.~A.}\ \bibnamefont
  {Kraut}}, \bibinfo {author} {\bibfnamefont {R.~W.}\ \bibnamefont {Grant}},
  \bibinfo {author} {\bibfnamefont {J.~R.}\ \bibnamefont {Waldrop}},\ and\
  \bibinfo {author} {\bibfnamefont {S.~P.}\ \bibnamefont {Kowalczyk}},\
  }\bibfield  {title} {\bibinfo {title} {{Semiconductor core-level to
  valence-band maximum binding-energy differences: Precise determination by
  x-ray photoelectron spectroscopy}},\ }\href
  {https://doi.org/10.1103/PhysRevB.28.1965} {\bibfield  {journal} {\bibinfo
  {journal} {Phys. Rev. B}\ }\textbf {\bibinfo {volume} {28}},\ \bibinfo
  {pages} {1965} (\bibinfo {year} {1983})}\BibitemShut {NoStop}%
\bibitem [{Note4()}]{Note4}%
  \BibitemOpen
  \bibinfo {note} {In this case we are limited to $\beta =10$ eV$^{-1}$ for
  computational reasons.}\BibitemShut {Stop}%
\bibitem [{\citenamefont {Rinke}\ \emph {et~al.}(2005)\citenamefont {Rinke},
  \citenamefont {Qteish}, \citenamefont {Neugebauer}, \citenamefont
  {Freysoldt},\ and\ \citenamefont {Scheffler}}]{Rinke2005}%
  \BibitemOpen
  \bibfield  {author} {\bibinfo {author} {\bibfnamefont {P.}~\bibnamefont
  {Rinke}}, \bibinfo {author} {\bibfnamefont {A.}~\bibnamefont {Qteish}},
  \bibinfo {author} {\bibfnamefont {J.}~\bibnamefont {Neugebauer}}, \bibinfo
  {author} {\bibfnamefont {C.}~\bibnamefont {Freysoldt}},\ and\ \bibinfo
  {author} {\bibfnamefont {M.}~\bibnamefont {Scheffler}},\ }\bibfield  {title}
  {\bibinfo {title} {{Combining {$GW$} calculations with exact-exchange
  density-functional theory: an analysis of valence-band photoemission for
  compound semiconductors}},\ }\href
  {https://doi.org/10.1088/1367-2630/7/1/126} {\bibfield  {journal} {\bibinfo
  {journal} {New Journal of Physics}\ }\textbf {\bibinfo {volume} {7}},\
  \bibinfo {pages} {126} (\bibinfo {year} {2005})}\BibitemShut {NoStop}%
\bibitem [{\citenamefont {Aryasetiawan}\ and\ \citenamefont
  {Gunnarsson}(1996)}]{Aryasetiawan1996}%
  \BibitemOpen
  \bibfield  {author} {\bibinfo {author} {\bibfnamefont {F.}~\bibnamefont
  {Aryasetiawan}}\ and\ \bibinfo {author} {\bibfnamefont {O.}~\bibnamefont
  {Gunnarsson}},\ }\bibfield  {title} {\bibinfo {title} {{$3d$ semicore states
  in ZnSe, GaAs, and Ge}},\ }\href {https://doi.org/10.1103/PhysRevB.54.17564}
  {\bibfield  {journal} {\bibinfo  {journal} {Phys. Rev. B}\ }\textbf {\bibinfo
  {volume} {54}},\ \bibinfo {pages} {17564} (\bibinfo {year}
  {1996})}\BibitemShut {NoStop}%
\bibitem [{\citenamefont {Rohlfing}\ \emph {et~al.}(1997)\citenamefont
  {Rohlfing}, \citenamefont {Kr\"uger},\ and\ \citenamefont
  {Pollmann}}]{Rohlfing1997}%
  \BibitemOpen
  \bibfield  {author} {\bibinfo {author} {\bibfnamefont {M.}~\bibnamefont
  {Rohlfing}}, \bibinfo {author} {\bibfnamefont {P.}~\bibnamefont {Kr\"uger}},\
  and\ \bibinfo {author} {\bibfnamefont {J.}~\bibnamefont {Pollmann}},\
  }\bibfield  {title} {\bibinfo {title} {{Quasiparticle calculations of
  semicore states in Si, Ge, and CdS}},\ }\href
  {https://doi.org/10.1103/PhysRevB.56.R7065} {\bibfield  {journal} {\bibinfo
  {journal} {Phys. Rev. B}\ }\textbf {\bibinfo {volume} {56}},\ \bibinfo
  {pages} {R7065} (\bibinfo {year} {1997})}\BibitemShut {NoStop}%
\bibitem [{\citenamefont {Miyake}\ \emph {et~al.}(2006)\citenamefont {Miyake},
  \citenamefont {Zhang}, \citenamefont {Cohen},\ and\ \citenamefont
  {Louie}}]{Miyake2006}%
  \BibitemOpen
  \bibfield  {author} {\bibinfo {author} {\bibfnamefont {T.}~\bibnamefont
  {Miyake}}, \bibinfo {author} {\bibfnamefont {P.}~\bibnamefont {Zhang}},
  \bibinfo {author} {\bibfnamefont {M.~L.}\ \bibnamefont {Cohen}},\ and\
  \bibinfo {author} {\bibfnamefont {S.~G.}\ \bibnamefont {Louie}},\ }\bibfield
  {title} {\bibinfo {title} {{Quasiparticle energy of semicore $d$ electrons in
  $\mathrm{ZnS}$: Combined $\mathrm{LDA}+U$ and {$GW$} approach}},\ }\href
  {https://doi.org/10.1103/PhysRevB.74.245213} {\bibfield  {journal} {\bibinfo
  {journal} {Phys. Rev. B}\ }\textbf {\bibinfo {volume} {74}},\ \bibinfo
  {pages} {245213} (\bibinfo {year} {2006})}\BibitemShut {NoStop}%
\end{thebibliography}%

\end{document}